\documentclass[sigconf]{acmart}

\AtBeginDocument{%
  }

\usepackage{url}
\usepackage{xspace}
\usepackage{nccmath} 
\usepackage{multirow}
\usepackage{booktabs}
\usepackage{bigstrut}
\usepackage{bm}
\usepackage{graphicx}
\usepackage{xcolor}
\usepackage{bbding}

\newcommand{\revise}[1]{\textcolor{black}{#1}}

\usepackage{algorithm}
\usepackage{algorithmic}

\setcopyright{rightsretained}
\copyrightyear{2024}
\acmYear{2024}
\acmDOI{10.1145/3626202.3637562}

\acmConference[FPGA '24]{Proceedings of the 2024 ACM/SIGDA International Symposium on Field Programmable Gate Arrays}{March 3--5, 2024}{Monterey, CA, USA}
\acmBooktitle{Proceedings of the 2024 ACM/SIGDA International Symposium on Field Programmable Gate Arrays (FPGA '24), March 3--5, 2024, Monterey, CA, USA}
\acmISBN{979-8-4007-0418-5/24/03}

\begin{document}

\title[FlightLLM: Efficient Large Language Model Inference with a Complete Mapping Flow on FPGAs]{FlightLLM: Efficient Large Language Model Inference\\with a Complete Mapping Flow on FPGAs}

\author{Shulin Zeng}
\authornote{Both authors contributed equally to this research.}
\affiliation{
  \institution{Tsinghua University}
  \institution{Infinigence-AI}
  \country{}
}
\author{Jun Liu}
\authornotemark[1]
\affiliation{
  \institution{Shanghai Jiao Tong University}
  \institution{Infinigence-AI}
  \country{}
}

\author{Guohao Dai}
\authornote{Corresponding authors. Email: daiguohao@sjtu.edu.cn, yu-wang@tsinghua.edu.cn}
\affiliation{
  \institution{Shanghai Jiao Tong University}
  \institution{Infinigence-AI}
  \country{}
}

\author{Xinhao Yang}
\affiliation{
  \institution{Tsinghua University}
  \institution{Infinigence-AI}
  \country{}
}

\author{Tianyu Fu}
\affiliation{
  \institution{Tsinghua University}
  \institution{Infinigence-AI}
  \country{}
}

\author{Hongyi Wang}
\affiliation{
  \institution{Tsinghua University}
  \institution{Infinigence-AI}
  \country{}
}

\author{Wenheng Ma}
\affiliation{
  \institution{Tsinghua University}
  \country{}
}

\author{Hanbo Sun}
\affiliation{
  \institution{Tsinghua University}
  \country{}
}

\author{Shiyao Li}
\affiliation{
  \institution{Tsinghua University}
  \institution{Infinigence-AI}
  \country{}
}

\author{Zixiao Huang}
\affiliation{
  \institution{Tsinghua University}
  \country{}
}

\author{Yadong Dai}
\affiliation{
  \institution{Infinigence-AI}
  \country{}
}

\author{Jintao Li}
\affiliation{
  \institution{Infinigence-AI}
  \country{}
}

\author{Zehao Wang}
\affiliation{
  \institution{Infinigence-AI}
  \country{}
}

\author{Ruoyu Zhang}
\affiliation{
  \institution{Infinigence-AI}
  \country{}
}

\author{Kairui Wen}
\affiliation{
  \institution{Infinigence-AI}
  \country{}
}

\author{Xuefei Ning}
\affiliation{
  \institution{Tsinghua University}
  \country{ }
}

\author{Yu Wang}
\authornotemark[2]
\affiliation{
  \institution{Tsinghua University}
  \country{ }
}

\renewcommand{\shortauthors}{Shulin Zeng and Jun Liu, et al.}

\begin{abstract}
Transformer-based Large Language Models (LLMs) have made a significant impact on various domains. However, LLMs' efficiency suffers from both heavy computation and memory overheads.
Compression techniques like sparsification and quantization are commonly used to mitigate the gap between LLM's computation/memory overheads and hardware capacity. However, existing GPU and transformer-based accelerators cannot efficiently process compressed LLMs, due to the following unresolved challenges: low computational efficiency, underutilized memory bandwidth, and large compilation overheads.

This paper proposes \textbf{FlightLLM}, enabling efficient LLMs inference with a complete mapping flow on FPGAs. 
In FlightLLM, we highlight an innovative solution that the computation and memory overhead of LLMs can be solved by utilizing FPGA-specific resources (\textit{e.g.}, DSP48 and heterogeneous memory hierarchy). 
We propose a configurable sparse DSP chain to support different sparsity patterns with high computation efficiency.
Second, we propose an always-on-chip decode scheme to boost memory bandwidth with mixed-precision support.
Finally, to make FlightLLM available for real-world LLMs, we propose a length adaptive compilation method to reduce the compilation overhead.  
Implemented on the Xilinx Alveo U280 FPGA, FlightLLM achieves 6.0$\times$ higher energy efficiency and 1.8$\times$ better cost efficiency against commercial GPUs (\textit{e.g.}, NVIDIA V100S) on modern LLMs (\textit{e.g.}, LLaMA2-7B) \revise{using vLLM and SmoothQuant under the batch size of one.} FlightLLM beats NVIDIA A100 GPU with 1.2$\times$ higher throughput using the latest Versal VHK158 FPGA.


\end{abstract}







\maketitle
\section{Introduction}

\begin{figure}[t]
  \centering
  \includegraphics[width=0.95\linewidth]{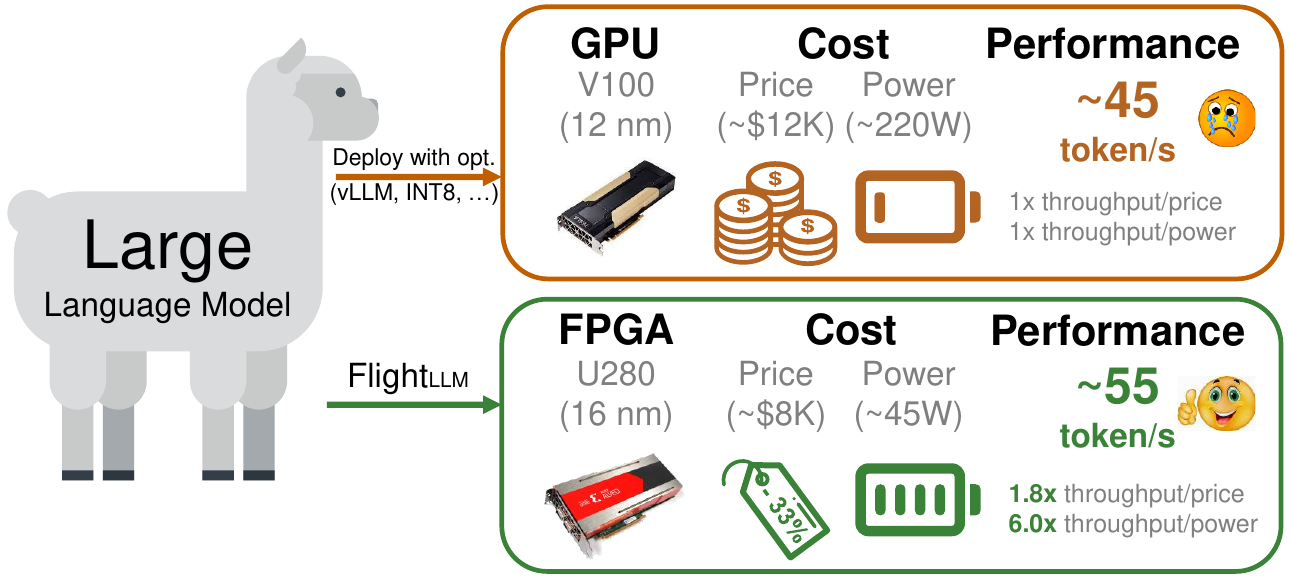}
  \vspace{-5pt}
  \caption{FlightLLM on Alveo U280 FPGA outperforms NVIDIA V100S GPU \revise{(using vLLM~\cite{kwon2023efficient} and SmoothQuant~\cite{xiao2023smoothquant})} with better performance and cost efficiency.}
  \vspace{-20pt}
  \label{fig:intro}
\end{figure}

Recently, we have witnessed the rapid development and significant impact of Large Language Models (LLMs)~\cite{bommasani2021opportunities, wei2022emergent}. 
LLMs demonstrate amazing power to understand all the users' input requests (\textit{prefill} stage) and generate accurate responses token-by-token (\textit{decode} stage).
\revise{LLMs are being widely used in latency-sensitive scenarios~\cite{Naveed2023ACO}, }
such as \revise{code completion~\cite{Wang2023PractitionersEO},} real-time chatbots~\cite{chen2023llm, thirunavukarasu2023large}, customer support~\cite{jeong2023study}, online legal advice~\cite{230616092}, and beyond. 
\revise{The latency is critical for a good user experience, and the batch size is usually set as 1 to meet the real-time requirement.}
However, current LLMs suffer from both heavy computation and memory overheads because of the explosive growth model size of LLMs. Taking GPT-3~\cite{brown2020language} as an example, it has 175 billion parameters (\textit{i.e.}, 350GB in FP16), requiring about 660TOPS of computation amount to complete a single inference.  
Model compression methods \cite{deng2020model} (\textit{e.g.}, sparsification, quantization, etc.) are commonly applied to address the above issues.
However, the unique computation schemes of these methods are not efficiently supported by current hardware platforms, like GPUs, for LLMs.
From the computation perspective, current GPUs only support structured sparsity (\textit{e.g.,} 2:4 sparsity), leading to significant algorithm accuracy loss of LLMs~\cite{frantar2023sparsegpt}. 
In contrast, the unstructured sparsity ensuring algorithm accuracy cannot bring end-to-end acceleration for LLMs. 
For example, the 75\% unstructured sparsity only leads to negligible end-to-end speedup~\cite{feng2023diffuser}. 
From the memory perspective, quantization and large on-chip memory can reduce data access. 
Recent algorithm studies~\citep{squeezellm,llmint8} are pushing the limit of bit-width with mixed-precision quantization. 
However, the alignment feature of GPU's cache and SIMD architecture requires homogeneous bit-widths of LLM parameters for weight access reduction~\cite{xiao2023smoothquant}. 
Compounding the issue, GPU's KB-scaled share memory of SMs cannot hold all the activations for LLM text generation.

\begin{figure}[t]
  \centering
  \includegraphics[width=0.95\linewidth]{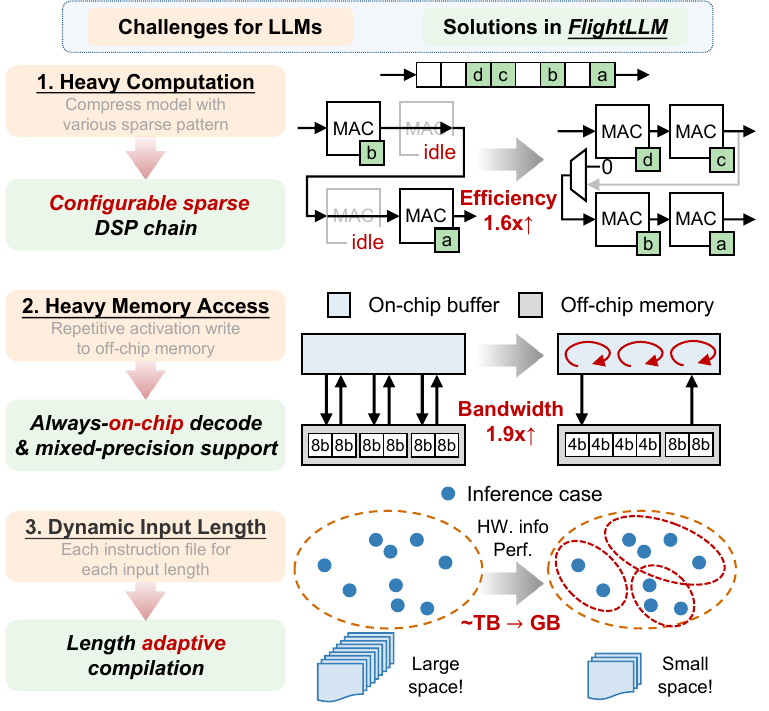}
  \vspace{-15pt}
  \caption{Three challenges of LLM inference on FPGAs, and the corresponding solutions in FlightLLM.}
  \vspace{-20pt}
  \label{fig:overview}
\end{figure}

FPGAs are potential solutions to accelerate LLM inference and explore the benefits brought by model compression, which has been proven in previous deep learning models~\cite{guo2017angel, wang2022logic, gong2022n3h, sun2022film, zhang2021fracbnn}. However, efficient LLM inference on FPGAs needs to solve the following challenges (Fig.~\ref{fig:overview}):
\begin{itemize}
    \item \textbf{Low computation efficiency.} Flexible sparsity patterns (\textit{e.g.,} block sparsity~\cite{zaheer2020bigbird}, N:M sparsity~\cite{chen2023dynamic_nm}, etc.) in LLM leads to low computation efficiency.
    \item \textbf{Underutilized memory bandwidth.} The \textit{decode} stage of LLM repetitively accesses fine-grained data from off-chip memory, leading to underutilized bandwidth (\revise{29-43\%}). 
    \item \textbf{Large compilation overheads.} The dynamic sparsity patterns and input lengths of LLMs constitute a large design space. For example, generating instructions for 2048 input token length results in $\sim$TB storage overhead on FPGAs.  
\end{itemize}

In this paper, we propose \textbf{FlightLLM}, enabling efficient LLMs inference with a complete mapping flow on FPGAs \revise{(Fig.~\ref{fig:intro})}. 
FlightLLM innovatively points out that the computation and memory overhead of LLMs can be solved by utilizing FPGA-specific resources (\textit{e.g.}, DSP48 and heterogeneous memory hierarchy). 
To address the challenges of low computation efficiency, FlightLLM exploits a configurable sparse DSP chain. We introduce a flexible cascaded DSP48 architecture to support different sparsity patterns with high computation efficiency (\textit{i.e.}, runtime DSP utilization). 
To tackle the underutilized memory bandwidth, FlightLLM proposes an always-on-chip decode scheme. Activations reside in the on-chip memory during the \textit{decode} stage with the support of mixed-precision quantization. 
To reduce the compilation overhead, FlightLLM proposes a length adaptive compilation method. Instructions for consecutive input token length are grouped, and the total storage overhead for instructions can be reduced.  

The main contributions of this paper are as follows.
\begin{itemize}
    \item We propose a configurable sparse DSP chain to support different sparsity patterns. FlightLLM improves the computation efficiency by 1.6$\times$ with block-wise and N:M sparsity.
    \item We propose an always-on-chip decode scheme with mixed-precision support. FlightLLM boosts the memory bandwidth from \revise{35.6\%} to 65.9\%.
    \item We propose a length adaptive compilation method to reduce the instruction storage overhead by 500$\times$ ($\sim$GB), enabling deploying real-world LLMs onto FPGAs.
\end{itemize}

We implement FlightLLM on the Xilinx Alveo U280 FPGA~\footnote{Artifact is available at: \url{https://zenodo.org/doi/10.5281/zenodo.10422477}}. Evaluated on the OPT-6.7B and LLaMA2-7B, FlightLLM achieves better end-to-end latency than NVIDIA V100S GPU \revise{using vLLM~\cite{kwon2023efficient} and SmoothQuant~\cite{xiao2023smoothquant} under the batch size of one}. Besides, FlightLLM outperforms NVIDIA V100S and A100 GPU with 6.0$\times$ and 4.2$\times$ higher energy efficiency, and 1.8$\times$ and 1.4$\times$ better cost efficiency on average, respectively. When evaluated on the latest Versal VHK158 FPGA, FlightLLM beats NVIDIA A100 with 1.2$\times$ higher throughput.

\section{Background and Related Work}
\subsection{Background}
\begin{figure}[!tp]
  \centering
  \includegraphics[width=\linewidth]{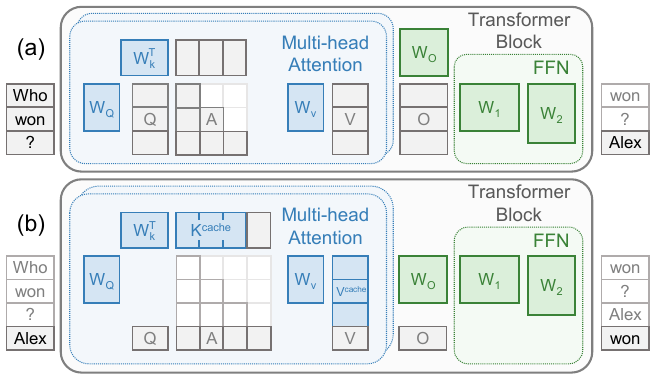}
  \vspace{-15pt}
  \caption{The (a) \textit{prefill} and (b) \textit{decode} stage of LLMs. Colored squares are weights or cached data. Gray denotes activations.}
  \vspace{-15pt}
  \label{fig:prefill-decode}
\end{figure}

Transformer-based LLMs~\citep{touvron2023llama2, zhang2205opt} achieve state-of-the-art (SOTA) performance across all kinds of Natural Language Processing (NLP) tasks. 
The transformer model architecture consists of many cascaded transformer blocks and each transformer block generally includes two types of networks: the Multi-Head Attention (MHA) and the Feed Forward Network (FFN). 

Given $N$ input tokens embedded in $D$ dimensional space $X \in \mathbb{R}^{N \times D}$, the MHA projects the token embedding as $h$ heads' query $Q$, key $K$ and value $V$ matrices in $\mathbb{R}^{h \times N \times D}$ space and performs attention operation for each head as shown in equation~\ref{eq:mha}.

\begin{equation}
    Q = XW_{Q}, K = XW_{K}, V = XW_{V};  \\
    O = \text{softmax}(QK^{T})V
\label{eq:mha}
\end{equation}
where $W_{Q}$, $W_{K}$, $W_{V}$ represent the weight matrix in MHA.

The FFN further transforms each token embedding. Given the MHA output matrix $O \in \mathbb{R}^{h \times N \times D}$, 
FFN passes it through two fully connected layers with a non-linear activation function $g$ and generates the token embedding for the next transformer block.

\begin{equation}
    X = g(O W_1)W_2
\label{eq:ffn}
\end{equation}
where, $W_{1}$, $W_{2}$ represent the two weight matrix in FFN.

As shown in Fig.~\ref{fig:prefill-decode}, the workflow of LLMs can be divided into two main stages: the prefill stage and the decode stage.
In the prefill stage, LLM takes a prompt from the user which is a sequence of tokens as the input (e.g. the "\textit{Who won ?}" in Figure.\ref{fig:prefill-decode} (a)).
Then, LLM will understand the context of the prompt and generates the first response token (e.g. the "\textit{Alex}" in Figure.\ref{fig:prefill-decode} (a)).
All the $N$ input tokens are processed simultaneously with high throughput.
In the decode stage, LLM treats the newly generated token as length $N=1$ input and generates the next token (e.g. the "\textit{won}" in Figure.\ref{fig:prefill-decode} (b)). Since LLM only processes one new token at a time in the decode stage, the matrix-matrix multiplications in equation~\ref{eq:mha} and ~\ref{eq:ffn} become matrix-vector multiplications.
The decode stage is called iteratively to generate the response token by token.

\vspace{-5pt}
\subsection{Related Work}
\textbf{Efficient Transformer.}
To tackle the extreme overhead of LLMs, various compression techniques are commonly used.
Quantization~\citep{dettmers2022llm_int8, frantar2022gptq, yao2022zeroquant, xiao2023smoothquant,lin2023awq} approaches use low-bit integers to substitute the 16-bit floating point parameters and activations for inference. 
Sparse attention~\citep{child2019spTrans, zaheer2020bigbird, beltagy2020longformer, wang2021spatten, chen2023dynamic_nm} and weight pruning~\citep{frantar2023sparsegpt, kwon2022fast} approaches skip part of the attention matrix or weight matrix computing according to the defined sparse pattern. Various sparse patterns are proposed for different tasks and transformers, including local diagonal pattern~\citep{beltagy2020longformer}, block sparse~\citep{child2019spTrans, zaheer2020bigbird, kitaev2020reformer, kwon2022fast}, N:M sparse pattern~\citep{chen2023dynamic_nm, frantar2023sparsegpt}, row-column skipping pattern~\citep{wang2021spatten}, unstructured pattern~\citep{frantar2023sparsegpt} and so on.

\textbf{LLM-related Accelerators.}
Previous work~\cite{li2020ftrans,ham2020a3,isca2021elsa,micro2021sanger,wang2021spatten,micro2022DFX,micro2022bufferfly,hpca23flat,isca2023fact,wang2023cosa} propose customized architecture design for transformer models. 
Some work~\cite{CTA2023, micro2021sanger,wang2021spatten,micro2022bufferfly,li2020ftrans,isca2021elsa} lay more emphasis on accelerating sparse attention.
They design specialized architectures to fully utilize the pre-defined static attention pattern~\cite{micro2022bufferfly,li2020ftrans} or dynamically generated attention pattern~\cite{isca2023fact,CTA2023,wang2021spatten,micro2021sanger}.
Recently, FACT~\cite{isca2023fact} points out the importance of compressing linear layers with mixed-precision quantization to help reduce latency.
However, these methods cannot accelerate the decode stage of LLMs since they mainly focus on the \textit{prefill} stage for discriminative models, like medium-sized BERT~\cite{devlin2019bert} model.
DFX~\cite{micro2022DFX} emphasizes the acceleration of the decode stage of LLMs. However, it lacks hardware support for model compression of LLMs, making it hard to further expand model size or maximum token size.


\section{Computing Architecture}

\begin{figure}[!tp]
  \centering
  \includegraphics[width=\linewidth]{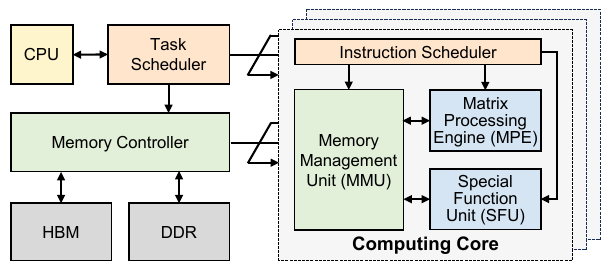}
  \vspace{-15pt}
  \caption{The overall architecture of FlightLLM, including task scheduler, memory controller and computing cores.}
  \vspace{-15pt}
  \label{fig:hw-overall}
\end{figure}

\subsection{Overall Architecture}
We design a high-performance FPGA-based accelerator for generative LLMs by making full use of FPGA resources. Combined with compression techniques like sparsification and quantization, FlightLLM can effectively accelerate the generative LLMs and reduce the inference overhead. 
As shown in Fig.~\ref{fig:hw-overall}, the overall hardware architecture of FlightLLM mainly includes a task scheduler, memory controller, and multiple computing cores (short as cores).
The accelerator uses model parallelism on multiple cores to complete the LLM inference task. The task scheduler assigns tasks to different cores and controls data synchronization. 

\begin{figure*}[!tp]
  \centering
  \includegraphics[width=\linewidth]{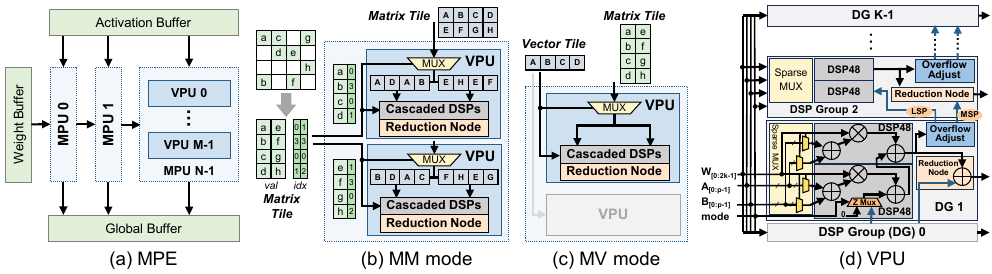}
  \vspace{-15pt}
  \caption{The unified Matrix Processing Engine (MPE) can perform multiple types of matrix multiplications. (a) MPE includes multiple Matrix Processing Units (MPUs), which are composed of multiple Vector Processing Units (VPUs). By configuring the MPU, the MPE can support both (b) matrix-matrix multiplication (MM) mode and (c) matrix-vector multiplication (MV) mode. (d) We utilize DSP resources on the FPGA to implement the VPU.}
  \vspace{-10pt}
  \label{fig:hw-mpe}
\end{figure*}

The components of each core include the unified Matrix Processing Engine (MPE), Memory Management Unit (MMU), Special Function Unit (SFU), and Instruction Scheduler. The instruction scheduler decodes the input instructions and schedules different hardware units to perform computations. The main functions of the remaining hardware units are as follows: \textbf{MPE} handles all matrix (\textit{i.e.}, dense and sparse) operations in LLMs. MPE utilizes the configurable sparse DSP chain to reduce the hardware overhead on FPGA.
\textbf{MMU} reduces memory access overheads by designing customized quantization units for low-bit mixed-precision and optimizing data placement for off-chip memory.
\textbf{SFU} handles miscellaneous operations (\textit{e.g.}, Softmax, etc.) besides matrix processing operations. It also provides an additional data path to share data with other SFUs in different cores, accelerating the MV operation.

\vspace{-5pt}
\subsection{Unified Matrix Processing Engine}

Although sparsification can bring huge theoretical benefits to LLM inference, they cannot directly achieve these benefits on existing architectures. To maximize the benefits of sparsification, we design the unified Matrix Processing Engine (MPE) to handle all operations related to matrix computation, including General Matrix Multiplication (GEMM), Sparse Matrix-Matrix multiplication (SpMM), General Matrix-Vector multiplication (GEMV), Sparse Matrix-Vector multiplication (SpMV), and Sampled-Dense-Dense Matrix Multiplication (SDDMM). As shown in Fig.~\ref{fig:hw-mpe}(a), the MPE includes multiple Matrix Processing Units (MPUs), which transfer weights from the weight buffer using the streaming approach. The activation buffer and the global buffer store the input and output activations of the MPE, respectively.
By configuring the MPU, the MPE can support both matrix-matrix multiplication (MM) (Fig.~\ref{fig:hw-mpe}(b)) and matrix-vector multiplication (MV) mode (Fig.~\ref{fig:hw-mpe}(c)). The MPU is composed of multiple vector processing units (VPUs). The VPU is the basic component in the MPE, which performs the dot product of two vectors.


We build the unified MPE to support all the five operator on the same hardware achitecture. FlightLLM overcomes the challenge of low computational efficiency through hardware/software co-design. 
To do this, we first introduce the MPU, which exploits configurable sparse DSP chain to reduce hardware overhead while supporting sparse reduction. We use the MM mode as an example to illustrate the main idea and the implementation of MPU. Then, we re-design MPE's parallel scheme to maximize the performance in the MV mode. Finally, we introduce the SDDMM support through simple instruction scheduling.


\vspace{-5pt}
\subsubsection{MPU Design}
In transformer-based LLMs, sparsification methods including sparse attention and weight pruning are widely used to accelerate the LLM inference. 
The sparse pruning generates sparse matrix, whose densities and sparse patterns are uncertain. It brings great challenges to hardware design, especially for FPGA-based architectures that use the fixed DSP48 as the multiplication unit. Existing work introduces large additional hardware architectures to support sparse computations, which leads to a significant increase in hardware resources (about 5$\times$~\cite{matraptor-2020}). Without proper architectural design, the benefit of sparsification is weakened.

We utilize the DSP48 engine on FPGA to support sparse operations.
In order to reduce the hardware overhead, previous work cascades the DSPs to take full advantage of the hardware resources in DSP48. DSP cascading makes the most use of the accumulator, the result carry-out port, and the result carry-in port, improving the hard-core utilization. However, the fully cascaded DSP architecture are not friendly to sparse computation since the cascaded chain is a fixed path. 
In FlightLLM, we propose a \textbf{configurable sparse DSP chain (CSD-Chain)} to supplement the fixed DSP chain. In the CSD-Chain, a long DSP chain is divided into several DSP groups. A DSP group (DG) has several DSP48 cores, that are cascaded in a fixed manner. \revise{We pack two INT8 MACs on DSP48~\cite{wp486}.} Different DGs are cascaded with a configurable path. A VPU is made up of a CSD-Chain and a MPU consists of several VPUs. 
Fig.~\ref{fig:hw-mpe}(d) shows the architecture of the CSD-Chain based VPU. Each DG has two DSP48 cores. We use configurable cascading to support sparse matrix operations by adding three units to the fixed DSP chain.

\textbf{Sparse Mux}. As shown in Fig.~\ref{fig:hw-mpe}(d), two activations (A and B) are delivered to one DSP48 core simultaneously for weight reuse. Before the delivery, they are sent to a sparse MUX unit. In the sparse MUX unit, each activation is selected from multiple inputs according to the sparse index (shown in Fig.~\ref{fig:hw-mpe}(b)). With this sparse-based multiplexer, only nonzero inputs are sent to the DSP48 core.

\begin{figure}[!tp]
  \centering
  \includegraphics[width=0.95\linewidth]{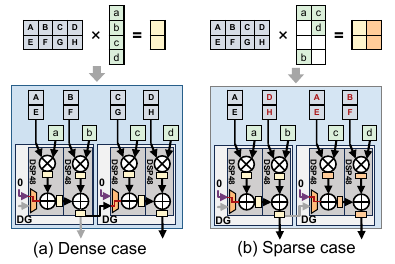}
  \vspace{-10pt}
  \caption{By configuring the VPU, the MPE can support both (a) dense and (b) sparse cases.}
  \vspace{-15pt}
  \label{fig:hw-vpu}
\end{figure}

\textbf{Reduction Node (RN)}. Compared to GEMM operations, calculating SpMM may produce more outputs, as demonstrated in Fig.~\ref{fig:hw-vpu}(b). Thus, DSPs are grouped in our design to implement non-breaking MAC and a reduction node are inserted at the end of a DG. When a SpMM operation wants to generate multiple outputs, the RN will break the configurable cascade path between DGs, and calculate the output. Other DGs on the CSD-Chain will start a new SpMM computation by selecting zero in the Z-MUX.

\textbf{Overflow Adjust Unit (OAU)}. When a DSP48 shares a weight with two activations, only 18 bits can be accessed for each activation. As a result, a long cascade accumulation may overflow. Therefore, we adjust the output data before sending it to the next DG. In the overflow adjust unit, the result is split into a most significant part (MSP) and a least significant part (LSP). 
The LSP cascades to the next DSP48 with limited bits to avoid accumulation overflow. The MSP are delivered to the RN in the next DG to calibrate the output result. 
In this way, all accumulators in DSP48 are fully utilized. Since a 18-bit integer will never overflow if no more than eight 16-bit integer are accumulated, the OAU is skipped with no more than eight DSP48 cores.

Due to the configurable cascade path, VPU with the CSD-Chain can efficiently work on both dense and sparse multiplications. 
As shown in Fig.~\ref{fig:hw-vpu}, all DSP48 cores are fully used in both cases. The only difference is that the RN in sparse case will break the CSD-Chain into two individual DSP chains to execute two different MACs and produce two outputs.

Supporting sparse matrix multiplication could improve the computation efficiency. But arbitrary sparsity may cause data mismatch between different DGs, leading to unexpected efficiency decrease. 
Existing work shows that N:M sparse pattern is a promising sparsification method. It maintains the same sparsity ratio within each matrix block, and allocates different sparsity ratios among different matrix blocks. Where M is an integer power of 2, and N is the partial factor of M. For example, M=16, N=0, 2, 4, 8, 16. The N:M sparse method restricts the number and position of nonzero elements while maintaining flexible sparsity. It can be easily mapped to a CSD-Chain. For a N:M sparse architecture, a CSD-Chain can be splited into N groups. Each DSP will select one input from M inputs. In each cycle, the entire CSD-Chain can produce one MAC output in dense case and N MAC outputs in N:M sparse case. Fig.~\ref{fig:hw-vpu} shows the case of a VPU supporting 2:4 sparse pattern.

\subsubsection{Matrix-Vector Multiplication Analysis}
We explore the hyper-parameter space of compute tiling to fully utilize the off-chip memory bandwidth.
We model the memory access time $T_{mem}$ and computing time $T_{cmp}$ of general MM in equation~\ref{eq:design_space}.
$M\times K$, $K\times N$, and $M\times N$ denote the shapes of two input matrices and one output matrix, respectively. $p_M$, $p_K$, and $p_N$ denote the three dimensions of computational parallelism in matrix computation. $BW$ denotes the off-chip bandwidth.

\begin{equation}
\begin{aligned}
T_{mem} &= \frac{M \cdot K + K \cdot N + M \cdot N}{BW} \\
T_{cmp} &= \frac{M \cdot K \cdot N}{p_M \cdot p_K \cdot p_N}
\label{eq:design_space}
\end{aligned}
\end{equation}

To overlap the computation and memory access with double-buffer, we need to make sure that $T_{mem} < T_{cmp}$.
For MV operations, $M=1$ and $p_M=1$ are set.
For the MV mode, we can iterate through the space to obtain a set of $[p_K', p_N']$ to guarantee the bound for double-buffer.
In other words, under the configuration of [$p_K'$, $p_N'$], we can realize that the MPE can still fully utilize the off-chip memory bandwidth in MV mode, although the computing resources in the MPE are partially idle at this time (Fig.~\ref{fig:hw-mpe}(c)). By redesigning the computational parallelism, MPUs can maximize the execution performance of executing GEMV and SpMV on FPGAs.

\subsubsection{SDDMM Support}
SDDMM is the key operator of the sparse self-attention layer. The block-wise sparsity of SDDMM can be used to reduce the amount of computation and improve the hardware energy efficiency. Therefore, we can treat SDDMM as multiple GEMMs in a block-wise manner. We only need to do some processing on the SDDMM operator with the instruction scheduler to efficiently complete the SDDMM computation on the MPE.


\begin{figure}[!tp]
  \centering
  \includegraphics[width=0.95\linewidth]{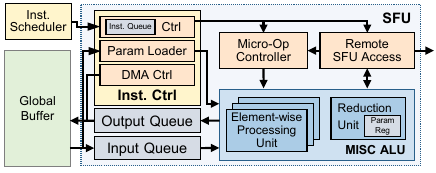}
  \vspace{-10pt}
  \caption{SFU contains a MISC ALU, an instruction controller, a micro-operator controller, and a remote SFU access engine.}
  \vspace{-15pt}
  \label{fig:hw-sfu}
\end{figure}

\vspace{-5pt}
\subsection{Special Function Unit}
Besides MM and MV computations, there are many other operations in LLMs, including softmax, layer normalization, etc. These miscellaneous (MISC) operations can be classified into two types: (a) Element-wise operation, which generates the result element by element (such as element-add and concat); (b) Two-phase operation, which will perform a reduction operation to get one or more parameters before the element-wise operation (such as softmax and layer normalization). Unlike MM and MV operations, these operations are not compute-intensive. Thus, we design a Special Function Unit (SFU) to handle all MISC operators, as shown in Fig.~\ref{fig:hw-sfu}. The SFU splits a MISC instruction into micro-operations, and delivers each micro-op to the ALU. The ALU will calculate the output according to the micro-op. All the input data is fetched from the MMU. For two-phase operations, the SFU will read an entire vector data from MMU to generate necessary parameters and read the same data again calculating the final output. For accuracy consideration,  softmax and layer normalization operations are calculated in fp16 since the hardware cost of SFU is acceptable. 

Hiding the computation latency of MISC operations is important to improve the end-to-end latency in LLMs. For MM and multi-head MV operations, MISC calculations can be hidden between different vectors. For MISC operations after single-head MV calculations, the SFU breaks the entire vector into several sub-vectors and performs MISC operations in fine granularity to hide the computation latency. In consideration of the scalability, multiple SFUs in different PEs may work together by accessing remote SFUs. A SFU can share parameters and calculation results with other SFUs. Thus, although a vector may be generated by different PEs simultaneously, the result could be sent to all other PEs without writing back to HBM. It reduces the end-to-end latency and the wire overhead on FPGA.

\section{Always-on-chip decode}

\subsection{On-chip Decode Dataflow}

In the decode stage, the main efficiency constraint arises from the frequent access of off-chip memory for small data-volume activation vectors.
To reduce off-chip memory access of activation vectors, we employ the concept of operator fusion and fuse the computation within each inference of decode stage. 
Consequently, we can significantly increase the off-chip HBM (High Bandwidth Memory) bandwidth utilization from \revise{about 35.6\%} to 65.9\%. 

Since the activations in the decode stage are small data-volume vectors instead of matrices, they can be fully accommodated by the on-chip buffer of the FPGA. 
Therefore, to reduce the frequent read and write operations of the activation vectors, we fuse the computations of all layers during each inference of the decode stage. 
Finally, the computation result is written back to the off-chip memory at the end of each inference of the decode stage. 

\begin{figure}[!tp]
    \centering
    \includegraphics[width=0.9\linewidth]{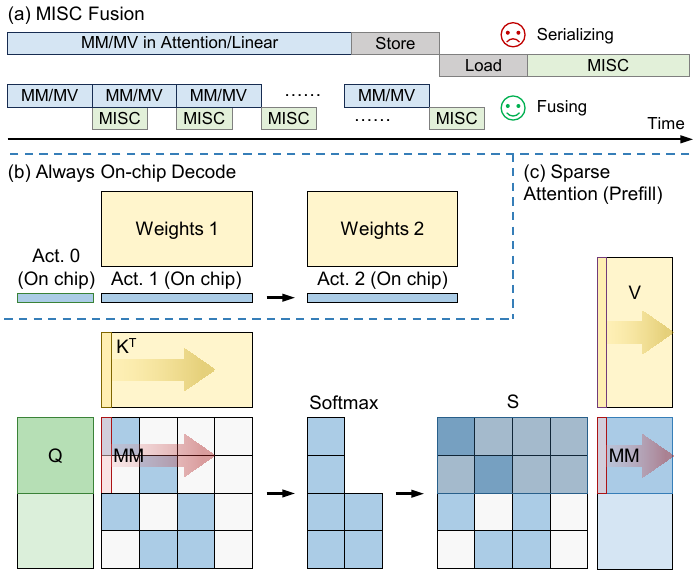}
    \vspace{-10pt}
    \caption{(a) MISC fusion with attention or linear operation. And an example of always on-chip decode approach in the (b) decode and (c) prefill stage.}
    \label{fig:dataflow-optimization}
    \vspace{-15pt}
\end{figure}

As depicted in Fig.~\ref{fig:dataflow-optimization}(b), given that we can directly use the output activation of the current layer as the input activation for the subsequent layer without writing the activation to off-chip memory, we retain the output activations from linear or attention operations within the on-chip buffer.
Through appropriate scheduling, the activation vector can consistently be stored in the on-chip buffer and then processed by either the MPE or SFU.

Furthermore, as illustrated in Fig.~\ref{fig:dataflow-optimization}(a), since there is no hardware resource conflict between SFU and MPE computations, we fuse the computations of these two different units to reduce the off-chip memory access of the intermediate results. 
Specifically, for the Softmax and LayerNorm operations, since they require a complete activation vector, it requires the \texttt{MV} operator in MPE to compute the activation vector before the \texttt{MISC} operator in SFU. For the activation functions (\textit{e.g.}, SiLU) and \revise{Element-wise addition/multiplication (Eltwise)}, they are \texttt{MISC} operators in SFU that can be computed immediately after the \texttt{MV} operators in MPE.

\vspace{-5pt}
\subsection{Discussion: Fusion in the Prefill}

The MISC fusion for the prefill stage is similar to that of the decode stage. However, the decode stage uses \texttt{MV} \revise{(see Table~\ref{tab:ISA} in Sec.~\ref{sec:ISA})}, while the prefill stage uses \texttt{MM} for matrix multiplication. Thus, in the prefill stage, Softmax and LayerNorm can start after the \texttt{MM} has processed an activation vector, while Eltwise and SiLU can start the computation after each \texttt{MM}.

In prefill attention computation, sparse attention can be applied. The sparse attention computation can be divided into 3 steps, computing $QK^T$, Softmax, and $SV$, as shown in Fig.~\ref{fig:dataflow-optimization}(c). 
When computing $QK^T$, the computation result is sparse according to the attention mask. 
When the zero attention mask completely covers the computation results, the corresponding \texttt{LD} and \texttt{MM} can be skipped. 
If the zero attention mask covers a part of the computation results, the \texttt{MM} instruction writes only the required part of the computation results back into the global buffer and then performs Softmax. 
When computing $SV$, the $S$ matrix is sparse, where the matrix $S$ stands for the attention matrix after the Softmax operation. 
Thus, in the proposed fused attention dataflow, Softmax and $SV$ are only computed for the parts that are not covered by the zero attention mask. The idea of acceleration by fusion is similar to the decode stage.

\vspace{-5pt}
\subsection{Mixed-precision Support}
The low-bit mixed-precision strategy can significantly reduce the parameters and the off-chip memory access for LLMs. However, GPU with the low-bit mixed-precision strategy (2/3/4/8-bit) makes it difficult to reduce memory access overhead. GPU uses SIMT-based computing architecture, which cannot efficiently process irregular and different bit-width data in memory. 
Therefore, we design a dedicated mixed-precision dequantization unit on the FPGA, which can efficiently process the compactly stored mixed-precision data in the buffer and convert it into a unified INT8 format to the MPE. \revise{Specifically, we transform mixed-precision multiplication (2/3/4-bit) into INT8 multiplication, to avoid excessive LUT overhead.} The dequantization unit consists of a set of parallel bit-width expansion units, which automatically expand the input data to 8 bits according to the control signal, scale factor, and sign bit.

\vspace{-5pt}
\subsection{Memory Latency Optimization}
We point out that the unique HBM+DDR hybrid memory system on FPGA has advantages over both HBM-only and DDR-only in reducing memory access overhead for generative LLMs. As we discussed in the previous section, the memory access patterns of SFU and MPE are significantly different. MPE needs to handle large-scale matrix multiplications, so single memory access is very large ($\sim$M Bytes). Using HBM can take full advantage of its high bandwidth. However, the operations processed by SFU include Eltwise, Softmax, etc., which are characterized by small single memory access data ($\sim$100 Bytes).
Because the memory access latency of HBM is higher than DDR, and the refresh cycle is less than DDR~\cite{zhao2018bandwidth,wang2020shuhai}. As a result, the overhead of HBM exceeds that of DDR when accessing a small amount of data. Therefore, we utilize the unique HBM+DDR hybrid memory system on the U280 FPGA to optimize the inference of the generative LLM. Specifically, we store small single-access data (\textit{e.g.}, Softmax, Silu, and Gelu lookup tables) on DDR to take advantage of the low memory access latency of DDR. We store large single-access data (\textit{e.g.}, KV cache, weights) on HBM to take advantage of the high memory bandwidth of HBM.

\section{Software Design}

\subsection{Instruction Set Architecture}\label{sec:ISA}

Instruction Set Architecture (ISA) acts as a connection between the LLM and hardware accelerator, consisting of six instructions listed in Table~\ref{tab:ISA}. \texttt{LD} and \texttt{ST} stand for transmission between off-chip (HBM or DDR) and on-chip memory. \texttt{MM} and \texttt{MV} represent for matrix-matrix and matrix-vector multiplication respectively. \texttt{MISC} controls other computations, including layer normalization (LayerNorm), SiLU, Softmax, and Eltwise. \texttt{SYS} is responsible for synchronization between multiple Super Logic Regions (SLRs) after each layer or with the host CPU after each inference is completed. 

\begin{table}[!tp]
    \caption{ISA design of FlightLLM.}\label{tab:ISA}
    \vspace{-10pt}
    \begin{tabular}{l|c}
    \toprule
    \textbf{Inst.} & \textbf{Discriptions} \\
    \midrule
    \texttt{LD}    & Load data from HBM or DDR to on-chip buffer. \\
    \texttt{ST}    & Store data from on-chip buffer to HBM or DDR. \\
    \texttt{MM}    & Calculate matrix-matrix multiplication $\textbf{C} = \textbf{XW}^T + \textbf{b}$. \\
    \texttt{MV}    & Calculate matrix-vector multiplication $\textbf{c} = \textbf{xW}^T + \textbf{b}$. \\
    \texttt{MISC}  & Calculate LayerNorm, SiLU, Softmax and Eltwise. \\
    \texttt{SYS}   & Synchronize between multiple SLRs or with host CPU. \\
    \bottomrule
    \end{tabular}
    \vspace{-10pt}
\end{table}

\vspace{-5pt}
\subsection{Length Adaptive Compilation}

\subsubsection{Challenges}
\revise{The instructions of FPGA accelerators are usually generated using static compilation, leading to a large instruction volume for different input shapes.}
Due to the fact that generative LLMs generate one token at a time, the token length increases by 1 with each inference. This means that each inference process of the LLM requires different instructions. However, due to the large computational and storage requirements of the LLM, even with coarse-grained instructions, the number of instructions is still enormous. Taking the example of deploying the LLaMA2-7B model on the U280 FPGA, the average volume of instructions required for the decode stage of each inference on each SLR is approximately 2.9 MB. For the prefill stage, the average volume of instructions required for each inference on each SLR is about 282.1 MB. Suppose we need to store instructions for all 3 SLRs, covering all token scenarios for prefill and decode 1-2048, in order to handle random input and output token quantities. In this case, the instructions would require approximately 1.67 TB. This size already far exceeds the capacity of U280 DDR. Unfortunately, due to sparse attention and N:M sparse pattern, each layer and each head in the LLM has a different sparse pattern, resulting in different instructions. Thus, it is not possible to reuse one set of instructions for multiple layers and attention heads. Therefore, we urgently need a method to reduce the size of the instruction sequence, allowing for the realization of inputs and outputs with arbitrary token lengths within limited storage.

\subsubsection{Solutions}
The fundamental reason for the large size of the instruction file is to adapt arbitrary lengths of prefill and decode stages, which requires storing instructions for all possible scenarios in memory. To address this issue, we can reuse the same instructions by allowing different lengths of prefill or decode. Specifically, by setting a threshold range, token lengths within this range can share the same instructions. For example, when the input token length is between 1 and 16, we can reuse the instructions for 16 tokens. Additionally, considering our N:M sparse block size (16$\times$16) and the size of the sparse attention block (64$\times$64), reusing instructions in this manner would not have a significant impact on performance.

\revise{We find that instructions are executed more frequently in the decode stage than in the prefill stage.} The bottleneck of the decode stage lies in memory access, which is directly proportional to the token length.
\revise{Therefore, we use more refined thresholds in the decode stage to avoid too much redundant computation.} Moreover, we can reuse the same instruction file by configuring different base memory addresses of PEs of different SLRs through registers. The instruction size can be reduced to 4.77 GB with these optimizations, which the DDR of U280 can already store.

\revise{We optimize the instructions for multiple HBM channels memory access to reduce the instruction size further.}
For example, in each PE, the A buffer and the global buffer are connected to 8 HBM channels, and each HBM channel requires an \texttt{LD} or \texttt{ST} instruction each time the data is moved between the buffer and the HBM. We combine these similar instructions into one instruction, and the hardware decoder decodes the single instruction into eight hardware instructions. The eight hardware instructions will be launched to eight HBM channels simultaneously, enabling the concurrent read/write of multiple HBM channels to utilize the HBM bandwidth fully. 
Through these optimizations, the instruction size is reduced to 3.25 GB and stored in the DDR memory with negligible performance loss.


\begin{figure}[!tp]
  \centering
  \includegraphics[width=\linewidth]{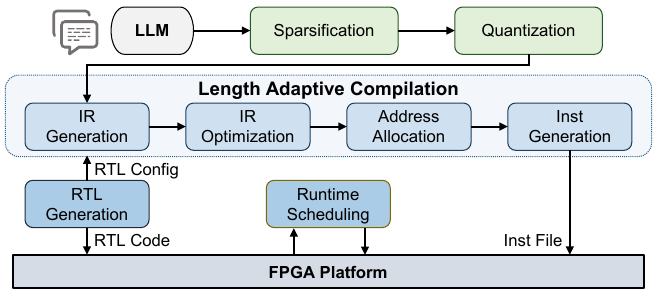}
  \vspace{-20pt}
  \caption{The overall mapping flow of FlightLLM.}
  \vspace{-15pt}
  \label{fig:compile_flow}
\end{figure}

\vspace{-5pt}
\subsection{Analytical Model for RTL Generation}
In this section, we analyze the relationship between the theoretical hardware resource utilization of FlightLLM on FPGA platforms and hardware implementation. For the main computation resource on FPGA, the usage of DSP is determined by the computing parallelism ($p_M * p_K * p_N$) of MPU and its amount. The theoretical usage of DSP can be estimated as follows: $ DSP = ( p_M * p_K * p_N * MPU) * MPE $. As for the on-chip buffers, their data width is designed based on the parallelism of the computation units. We implement activation buffer with URAM for its larger capacity, while the remaining buffers are implemented with BRAM36. The theoretical buffer usage are as follows: $URAM = ( p_M * p_K * Activation\_width/URAM\_width) * MPU * MPE$, $ BRAM = ( Weight\_buffer + Global\_buffer + \\Index\_buffer ) * MPE$.
For memory bandwidth, the theoretical peak bandwidth is calculated as $Bandwidth = ( MPU/8 + 2) * MPE * 14.4GB/s$.
The RTL generator determines the deployment of hardware modules of FlightLLM on a specific FPGA platform based on these theoretical models. It aims to fully utilize on-chip resources to improve the performance of the accelerator.

\vspace{-5pt}
\subsection{Mapping Flow}

Fig.~\ref{fig:compile_flow} shows our entire deployment flow. \revise{FlightLLM takes the PyTorch-based LLM as input and converts it to ISA according to the customized intermediate representation (IR).}
First, the original LLM undergoes sparsification and quantization to create a compressed LLM. Subsequently, the IR is exported, encompassing the model's structure, weights, sparse indexes, and attention masks, which is achieved through automated parsing of the model's structure. Following that, the generated IR undergoes optimization, which involves operations like removing the \texttt{view()} layers that do not impact the data arrangement and performing layer fusion. More specifically, the attention layer will be fused with the softmax layer, and the linear layer will be fused with ReLU, SiLU, and element-wise layers. Subsequently, all the data in the optimized IR will be assigned HBM or DDR storage addresses. Additionally, the data stored in the HBM will be partitioned into appropriate channels to prevent inefficient access across different channels, thus leveraging the FPGA's high bandwidth effectively. Lastly, the compiler will generate instructions using the optimized IR and schedule the on-chip buffer according to manually defined templates.

In addition, we support generating corresponding RTL for different FPGA platforms. Specifically, the RTL Generator takes parameters of different FPGA platforms (including the amount of DSP, the capacity and bandwidth of HBM/DDR and on-chip RAM resources) to dynamically adjust the computing parallelism and buffer size. This is done to generate corresponding RTL code for implementation and configurations for compilation, in order to maximize the optimal performance on different platforms.

\section{Evaluation}

\begin{figure}[!tp]
  \centering
  \includegraphics[width=0.8\linewidth]{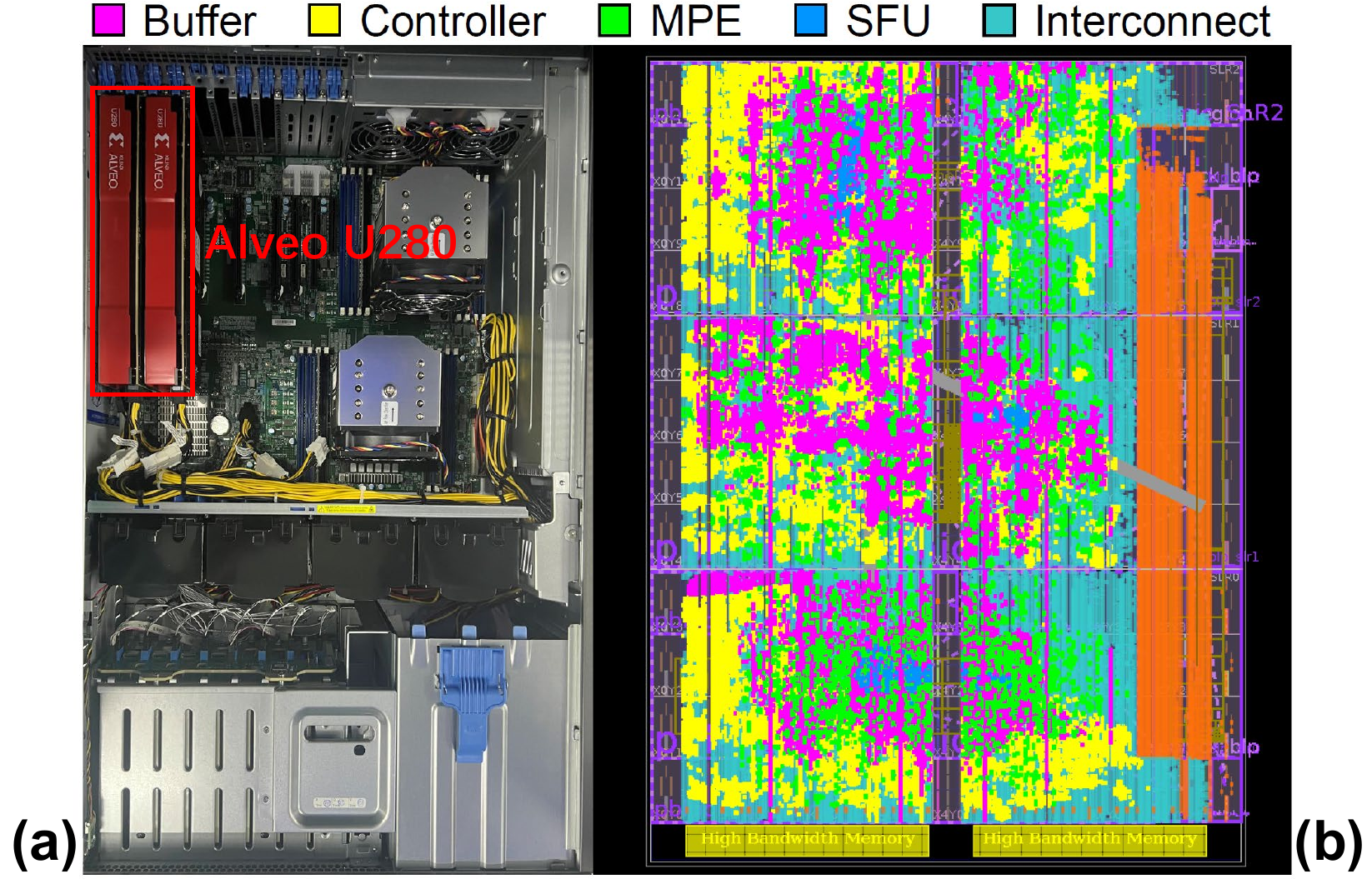}
  \vspace{-10pt}
  \caption{(a) U280 FPGAs on the server (one card is used only). (b) FlightLLM implementation layout on U280 FPGA.}
  \vspace{-15pt}
  \label{fig:exp-layer_out}
\end{figure}

\begin{table}[t]\footnotesize
\setlength\tabcolsep{4pt} 
\caption{Hardware parameters of GPU and FPGA platforms.}
\vspace{-10pt}
\begin{tabular}{lcccc}
\toprule
    & \textbf{GPU}  & \textbf{GPU}  & \textbf{FPGA}   & \textbf{FPGA}   \\ \midrule
\textbf{Platform}       & \begin{tabular}[c]{@{}c@{}}NVIDIA\\ V100S(12nm)\end{tabular}       & \begin{tabular}[c]{@{}c@{}}NVIDIA\\ A100(7nm)\end{tabular}       & \begin{tabular}[c]{@{}c@{}}Xilinx Alveo\\ U280(16nm)\end{tabular} & \begin{tabular}[c]{@{}c@{}}Xilinx Versal\\ VHK158(7nm)\end{tabular} \\ \midrule
\textbf{Frequency} & 1245 MHz   & 1065 MHz     & 225 MHz    & 225 MHz   \\ \midrule
\begin{tabular}[l]{@{}l@{}}\textbf{Computing} \\ \textbf{Units}\end{tabular}  & \begin{tabular}[c]{@{}c@{}}640 \\ Tensor Cores\end{tabular} & \begin{tabular}[c]{@{}c@{}}432 \\ Tensor Cores\end{tabular} & \begin{tabular}[c]{@{}c@{}}9024 \\  DSPs\end{tabular}  & \begin{tabular}[c]{@{}c@{}}7392 \\  DSPs\end{tabular} \\ \midrule
\textbf{Memory}    & 32 GB & 80 GB  & 8 \& 32 GB  & 32 \& 32 GB  \\ \midrule
\textbf{Bandwidth} & 1134 GB/s & 1935 GB/s & 460 \& 38 GB/s & 819 \& 51 GB/s \\ \bottomrule
\vspace{-15pt}
\end{tabular}
\label{tab:exp_hardware_platforms}
\end{table}

\begin{table}[t]
\footnotesize
\setlength\tabcolsep{3pt}
\caption{Hardware utilization of FlightLLM on Alveo U280.}
\vspace{-10pt}
\begin{tabular}{lccccc}
\toprule
\textbf{Component}                                                   & \textbf{LUT}  & \textbf{FF}  & \textbf{BRAM} & \textbf{URAM} & \textbf{DSP}  \\ 
\midrule
\textbf{Buffer}                                                      & 42k(3.2\%)    & 75k(2.9\%)   & 816(40.5\%)   & 792(82.5\%)   & 0             \\ \midrule
\textbf{Controller} & 162k(12.4\%)  & 156k(6.0\%)  & 408(20.2\%)   & 0             & 0             \\ \midrule
\textbf{MPE}                                                     & 190k(14.6\%) & 360k(13.8\%) & 0             & 0             & 6144(68.1\%)  \\ \midrule
\textbf{SFU}                                                   & 30k(2.3\%)    & 36k(1.4\%)   & 24(1.2\%)     & 0             & 201(2.1\%)    \\ \midrule
\textbf{Interconnect}                                                & 150k(11.5\%)  & 316k(12.1\%) & 4(0.2\%)      & 0             & 0             \\ \midrule
\textbf{Total}                                                       & 574k(44.0\%)  & 943k(36.2\%) & 1252(62.1\%)  & 792(82.5\%)   & 6345(70.2\%) \\
\bottomrule
\vspace{-15pt}
\end{tabular}
\label{tab:exp_utilization}
\end{table}

\subsection{Evaluation Setup}

\noindent \textbf{Models and Datasets.}
We evaluate the effectiveness of FlightLLM with state-of-the-art large language models LLaMA2-7B~\citep{touvron2023llama2} and OPT-6.7B~\citep{zhang2205opt}. 
We finetune the compressed model with a small sampled subset of RedPajama dataset~\cite{together2023redpajama} consisting 8192 rows with 56M tokens.
The accuracy evaluation is performed on the commonly used
WikiText-103 and WikiText-2~\citep{merity2016wikitext2} datasets.

\noindent \textbf{Metrics.}
We leverage latency and throughput to evaluate the difference between FlightLLM and other baselines comprehensively. Latency is used to measure the end-to-end time cost of the entire inference. Throughput is used to measure the speed of the decode stage by dividing the number of output tokens by the time of the decode stage. \revise{Unless otherwise noted, all results are evaluated under the batch size of 1 to accommodate latency-sensitive scenarios.}

\noindent \textbf{FPGA Platforms.}
We use two FPGA platforms for evaluation, including Xilinx Alveo U280~\cite{fpgaU280} and Versal VHK158~\cite{VersalVH1582}. Table~\ref{tab:exp_hardware_platforms} lists the hardware parameters of FPGA platforms. Alveo U280 FPGA is equipped with two kinds of memory: 8GB HBM with 460GB/s bandwidth and 32GB DDR with 38GB/s bandwidth. Versal VHK158 FPGA's HBM capacity and bandwidth are significantly improved compared to U280, reaching 32GB and 819GB/s. We implement FlightLLM on the real system with U280 FPGAs (Fig.~\ref{fig:exp-layer_out}(a)). \revise{For VHK158 evaluation, we implement a cycle-accurate simulator, which has been verified with RTL emulation using Vitis 2023.1.}

\begin{figure*}[t]
  \centering
  \includegraphics[width=0.90\linewidth]{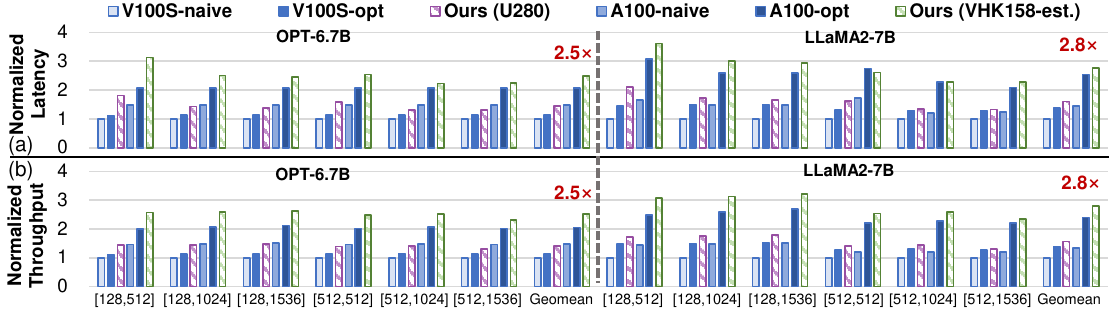}
  \vspace{-10pt}
  \caption{Latency and throughput of FlightLLM and V100S/A100 GPU. The horizontal axis represents [prefill size, decode size].}
  \vspace{-10pt}
  \label{fig:exp-perf-gpu}
\end{figure*}

\noindent \textbf{FPGA Implementation.}
Fig.~\ref{fig:exp-layer_out} shows the layout of our implementation on U280 FPGA. Since cross-die connections are more likely to become the critical paths for timing closure, we instantiate multiple computing cores and place them on different SLRs. For the memory controller, we place it on SLR0 closest to the HBM for easy reading. The implementation report shows that our kernel runs at 225 MHz, while detailed hardware utilization is listed in Table~\ref{tab:exp_utilization}. We also measure the power of FlightLLM through the vendor-provided Xilinx Board Utility tool xbutil~\cite{xbutil}.

\noindent \textbf{GPU Baselines.}
We choose NVIDIA A100 and V100S as our GPU baselines, and their specifications are also listed in Table~\ref{tab:exp_hardware_platforms}. We conduct evaluations on the selected model with huggingface PyTorch as the GPU-\textit{naive} design, vLLM~\cite{kwon2023efficient}, and SmoothQuant~\cite{xiao2023smoothquant} as the GPU-\textit{opt} design. vLLM is the commonly used LLM framework with KV cache memory optimization, and SmoothQuant is the SOTA LLM quantization framework with INT8 CUDA kernel for both activations and weights. 
We use nvprof~\cite{NvProf} to profile the GPU power consumption at runtime.

\noindent \textbf{SOTA Accelerator Baselines.} 
We also selected three domain-specific accelerators targeting at accelerating attention mechanism: DFX~\cite{micro2022DFX}, FACT~\cite{isca2023fact}, and CTA~\cite{CTA2023}. 
It has to be mentioned that DFX is a multi-FPGA acceleration work, and we only evaluate its hardware performance of a single card. 
Since there are no open-source codes for these accelerators, and they have not supported recent LLMs such as LLaMA2. We build C++ simulators based on corresponding hardware designs to evaluate their performance, \revise{ achieving less than 5\% deviation using their original data. For fairness, we align the hardware parameters (clock frequency, peak performance, bandwidth) for these baselines.}

\begin{table}[]
    \footnotesize
    \centering
    \caption{Perplexity of LLMs under different optimization configuration on wikitext-103 and wikitext-2 datasets.}
    \vspace{-10pt}
    \begin{tabular}{c|l|cc}
    \toprule
    \textbf{LLM} & \textbf{Compression} & \textbf{wikitext-103} & \textbf{wikitext-2}\\
    \midrule
    \multirow{5}{*}{\textbf{LLaMA2-7B}}  & None & 8.7 & 21.2 \\
    & Sparse Attention & 8.1 & 19.0 \\
    & Weight Pruning & 8.3 & 17.8 \\
    & Quantization & 9.9 & \revise{20.6} \\
    & All & 10.2 & 21.9 \\
    \midrule
    \multirow{5}{*}{\textbf{OPT-6.7B}}  & None & 11.0 & 10.0 \\
    & Sparse Attention & 11.1 & 10.5 \\
    & Weight Pruning & 11.8 & 11.1 \\
    & Quantization & 10.8 & 10.3 \\
    & All & 13.0 & 12.5 \\
    \bottomrule
    \end{tabular}
    \vspace{-15pt}
    \label{tab:evaluation_results}
\end{table}


\vspace{-5pt}
\subsection{Evaluation Results}
\subsubsection{Accuracy of Compressed LLMs}
FlightLLM harnesses the power of state-of-the-art model compression methods by optimizing them to fit the distinct features of FPGA. 
As depicted in Table~\ref{tab:evaluation_results}, we conduct experiments around different optimization configurations on LLMs.
For sparsification, 
FlightLLM builds upon previous work to use block sparse for sparse attention~\citep{zaheer2020bigbird} and N:M sparse for weight pruning~\cite{nm-sparse-arbitrary}. 
FlightLLM uses gradient-based analysis~\citep{dettmers2022llm_int8} to quantify the importance of each weight and attention value and remove the unimportant values.
For quantization, FlightLLM builds upon previous work~\cite{xiao2023smoothquant} and extends its single-precision scheme to mix-precision. 
FlightLLM follows the same idea as sparsification and use the gradiant-based analysis to quantify weight importance and assign three, four or five bit width accordingly. 
With this scheme, FlightLLM achieves average 3.5-bit for weights and 8-bit for activations.
\revise{Note the compression methods are also compatible with GPU, but they need the customized units of FPGA to yield real wall-clock speedup.}
By using the block sparse attention, N:M weight pruning and mixed-precision quantization all together, FlightLLM successfully compresses the original LLM with minimum perplexity influence.


\begin{table}[!tp]
    \centering
    \caption{\revise{The bandwidth utilization of different platform.}}
    \vspace{-10pt}
    \resizebox{0.8\columnwidth}{!}{
    \begin{tabular}{c|cc|cc|c|c}
    \toprule
    \textbf{Platform}              & \multicolumn{2}{c|}{V100S GPU}   & \multicolumn{2}{c|}{A100 GPU}    & U280             & VHK158           \\ \midrule
    \textbf{Solution}          & None    & Opt.    & None    & Opt.    & Ours        & Ours        \\ \midrule
    \textbf{BW Util.} & 42.5\% & 65.5\%  & 28.6\% & 57.4\% & \textbf{65.9\%} & \textbf{64.8\%} \\ \bottomrule
    \end{tabular}
    }
    \label{tab:BW-utilization}
    \vspace{-15pt}
\end{table}


\vspace{-5pt}

\subsubsection{Comparison with GPUs}
We compare the end-to-end latency of GPUs and FlightLLM. Fig.~\ref{fig:exp-perf-gpu} shows that FlightLLM on VHK158 outperforms V100S and A100 GPU in latency on both models with different combinations of input token size and output token size. For OPT-6.7B/LLaMA2-7B model, FlightLLM on U280 improves the end-to-end latency by $1.5/1.6\times$ and $1.3/1.2\times$ compared to V100S-\textit{naive} and V100S-\textit{opt}, respectively. 
\revise{Table~\ref{tab:BW-utilization} shows the bandwidth utilization of FlightLLM on FPGAs is better than A100 GPUs.}
This is because FlightLLM can be customized to design hardware units, which can fully exploit the sparsity of LLM and the memory access optimization in the decode stage.

\begin{figure*}[!tp]
  \centering
  \includegraphics[width=0.90\linewidth]{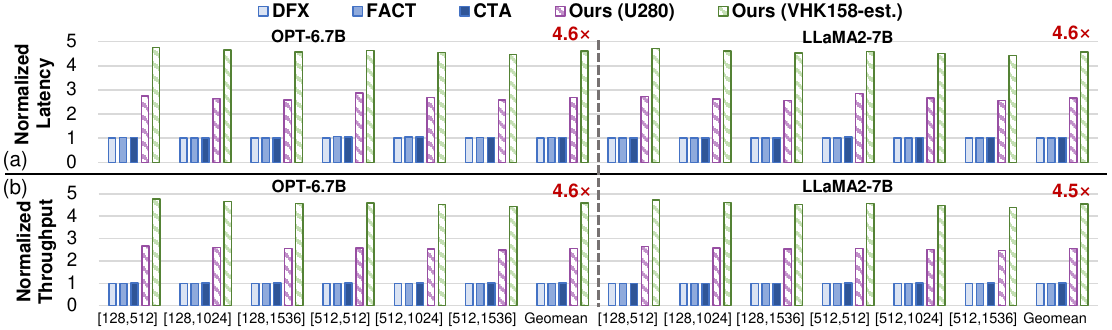}
  \vspace{-10pt}
  \caption{Performance of FlightLLM, DFX, CTA, and FACT. The horizontal axis represents [prefill size, decode size].}
  \vspace{-10pt}
  \label{fig:exp-lat-acc}
\end{figure*}
\begin{figure*}[!tp]
  \centering
  \includegraphics[width=0.90\linewidth]{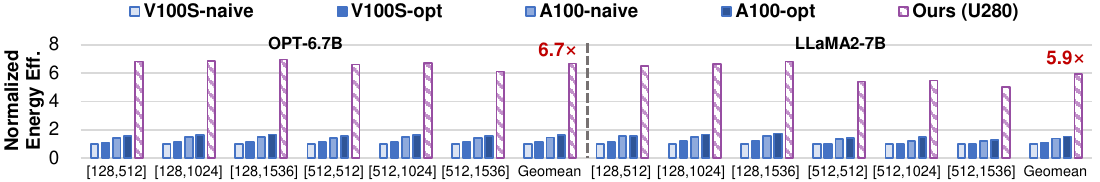}
  \vspace{-10pt}
  \caption{Energy efficiency of FlightLLM, NVIDIA V100S/A100 GPU. The horizontal axis represents [prefill size, decode size].}
  \vspace{-10pt}
  \label{fig:energy}
\end{figure*}
\vspace{-5pt}
\subsubsection{Comparison with SOTA accelerators}
Fig.~\ref{fig:exp-lat-acc}(a) shows the latency of different architectures running on OPT-6.7B and LLaMA2-7B model. 
It can be seen that FlightLLM achieves a general speed-up in end-to-end latency compared to DFX, CTA and FACT. The geomean latency speedups of FlightLLM on U280 and on VHK158 are 2.7$\times$ and 4.6$\times$ compared to DFX for OPT-6.7B, respectively. And the geomean throughput speedups of FlightLLM on U280 and on VHK158 are 2.6$\times$ and 4.6$\times$ compared to DFX. 
Compared to DFX, the acceleration effects of sparse attention in CTA and FACT are not significant, mainly because the attention computation does not account for a high proportion under small prefill size. However, our work adopts lower bit-width quantization scheme, which effectively alleviates the memory bottleneck in the decode stage. Fig.~\ref{fig:exp-lat-acc}(b) shows the geomean throughput of different architectures, with FlightLLM achieving the highest performance. Under the same hardware parameters, FlightLLM achieves better utilization of computing resources as well as bandwidth.

\begin{figure}[!tp]
  \centering
  \includegraphics[width=0.8\linewidth]{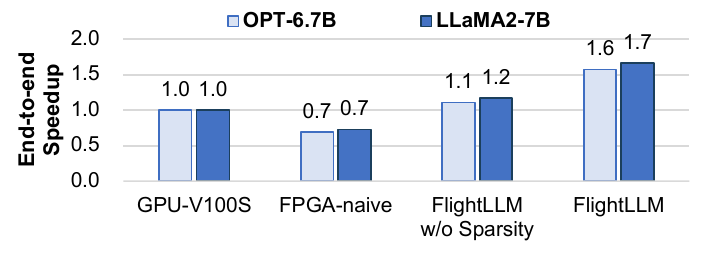}
  \vspace{-10pt}
  \caption{The latency breakdown of FlightLLM.}
  \vspace{-10pt}
  \label{fig:exp-perf-break}
\end{figure}
\begin{figure}[!tp]
  \centering
  \includegraphics[width=0.85\linewidth]{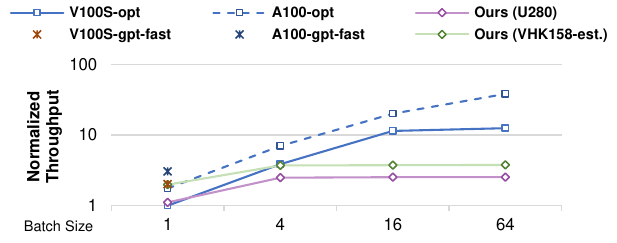}
  \vspace{-10pt}
  \caption{\revise{The multi-batch performance on LLaMA2-7B.}}
  \vspace{-18pt}
  \label{fig:exp-perf-batch}
\end{figure}

\vspace{-5pt}
\subsubsection{Energy and Cost Efficiency}
We consider energy efficiency as fair metrics to compare GPU and our FPGA-based FlightLLM. Fig.~\ref{fig:energy} shows the results of energy efficiency (Token/J). FlightLLM on U280 consistently outperforms GPUs and achieves $6.7\times$, $4.6\times$, $6.0\times$ and $4.2\times$ energy efficiency compared to V100S-\textit{naive}, A100-\textit{naive}, V100S-\textit{opt} and A100-\textit{opt} respectively for OPT-6.7B. For LLaMA2-7B, FlightLLM on U280 achieves achieves $6.0\times$, $4.4\times$, $5.5\times$ and $3.8\times$ energy efficiency.
For cost efficiency (Token/s/dollor), GPU generally has higher product price compared to FPGA. The price of V100S, A100 and Aleveo U280 are approximately 12000\$, 17000\$ and 8000\$ respectively. Therefore, FlightLLM on U280 achieves 1.9$\times$ and 1.5$\times$ higher geomean cost efficiency over V100S-\textit{opt} and A100-\textit{opt} for OPT-6.7B, and achieves 2.3$\times$ and 1.4$\times$ higher geomean cost efficiency over V100S-\textit{opt} and A100-\textit{opt} for LLaMA2-7B.

\vspace{-5pt}
\subsubsection{Performance Breakdown}
Fig.~\ref{fig:exp-perf-break} shows the latency breakdown of FlightLLM. We normalize the latency of the LLaMA-2 and OPT model running on a V100S GPU. We naively implement the LLaMA-2 model on the U280 FPGA, which has only 70\% of the performance of the V100S GPU. The gap is due to the larger peak memory bandwidth (1134GB/s vs. 460GB/s) and higher peak performance (130TOPS vs. 25TOPS) of the V100S GPU~\cite{a100-2021} compared to the U280 FPGA. After using the flexible sparse method and the configurable sparse DSP chain, the performance of FlightLLM is improved by 1.1-1.2$\times$, because we reduce the inference computation and make full use of DSP resources. After further using the always on-chip decoder, the performance improvement of FlightLLM achieves 1.6-1.7$\times$, because we effectively reduce the overhead of off-chip memory access.
\vspace{-5pt}
\subsubsection{\revise{Discussion}}
\revise{gpt-fast\footnote{\url{https://github.com/pytorch-labs/gpt-fast}, released in 2023.11.30.} is a new SOTA Pytorch-native codebase optimized for LLM inference, achieving 196.8 tokens/s with INT4 quantization on the NVIDIA A100 GPU. However, the current version of gpt-fast has no support for OPT models and multi-batch processing. Evaluated on LLaMA2-7B, FlightLLM on the VHK158 FPGA achieves 92.5 tokens/s, and provides 2.9$\times$ better energy efficiency and higher bandwidth utilization (64.8\% vs. 44.6\%) than gpt-fast on the A100 GPU.}

\revise{As for the multi-batch performance, gpt-fast has no support of multi-batch processing, so only the results of GPU-\textit{opt} (\textit{i.e.}, vllm and SmoothQuant) are reported. In Fig.~\ref{fig:exp-perf-batch}, as the batch size increases, the performance advantage of FlightLLM over GPU will gradually decrease. The main reason is that GPUs have more hardware resources (\textit{i.e.}, memory bandwidth and computing units with higher frequency) than FPGAs.}

\vspace{-5pt}
\section{Conclusion}
This paper proposes FlightLLM, enabling efficient LLMs inference with a complete mapping flow on FPGAs. 
In FlightLLM, we innovatively point out that the computation and memory overhead of LLMs can be solved by utilizing FPGA-specific resources. FlightLLM demonstrates that FPGAs are promising candidates for efficient LLM inference. FlightLLM achieves 6.0$\times$ higher energy efficiency and 1.8$\times$ better cost efficiency against commercial GPUs (\textit{e.g.}, NVIDIA V100S) on modern LLMs (\textit{e.g.}, LLaMA2).

\vspace{-5pt}
\begin{acks}
This work was supported by National Natural Science Foundation of China (No. 62325405, 62104128, U19B2019, U21B2031, 61832007, 62204164), and Beijing National Research Center for Information Science and Technology (BNRist). We thank for the insightful discussion with Shengen Yan and colleagues at Infinigence-AI. This work is supported by dgSPARSE project\footnote{The open source dgSPARSE project: \url{https://dgsparse.github.io/}}.
\end{acks}

\bibliographystyle{ACM-Reference-Format}
\balance
\bibliography{sample-base}


\begin{thebibliography}{57}


\ifx \showCODEN    \undefined \def \showCODEN     #1{\unskip}     \fi
\ifx \showDOI      \undefined \def \showDOI       #1{#1}\fi
\ifx \showISBNx    \undefined \def \showISBNx     #1{\unskip}     \fi
\ifx \showISBNxiii \undefined \def \showISBNxiii  #1{\unskip}     \fi
\ifx \showISSN     \undefined \def \showISSN      #1{\unskip}     \fi
\ifx \showLCCN     \undefined \def \showLCCN      #1{\unskip}     \fi
\ifx \shownote     \undefined \def \shownote      #1{#1}          \fi
\ifx \showarticletitle \undefined \def \showarticletitle #1{#1}   \fi
\ifx \showURL      \undefined \def \showURL       {\relax}        \fi
\providecommand\bibfield[2]{#2}
\providecommand\bibinfo[2]{#2}
\providecommand\natexlab[1]{#1}
\providecommand\showeprint[2][]{arXiv:#2}

\bibitem[wp4(2017)]%
        {wp486}
 \bibinfo{year}{2017}\natexlab{}.
\newblock \bibinfo{title}{Deep Learning with INT8 Optimization on Xilinx Devices}.
\newblock \bibinfo{howpublished}{[Online]}.
\newblock
\newblock
\shownote{\url{https://docs.xilinx.com/v/u/en-US/wp486-deep-learning-int8}}.


\bibitem[NvP(2022)]%
        {NvProf}
 \bibinfo{year}{2022}\natexlab{}.
\newblock \bibinfo{title}{nvprof}.
\newblock \bibinfo{howpublished}{[Online]}.
\newblock
\newblock
\shownote{\url{https://docs.nvidia.com/cuda/profiler-users-guide/index.html}}.


\bibitem[xbu(2022)]%
        {xbutil}
 \bibinfo{year}{2022}\natexlab{}.
\newblock \bibinfo{title}{Xilinx Board Utility Tool}.
\newblock \bibinfo{howpublished}{[Online]}.
\newblock
\newblock
\shownote{\url{https://xilinx.github.io/XRT/2021.1/html/xbutil2.html}}.


\bibitem[Beltagy et~al\mbox{.}(2020)]%
        {beltagy2020longformer}
\bibfield{author}{\bibinfo{person}{Iz Beltagy}, \bibinfo{person}{Matthew~E Peters}, {and} \bibinfo{person}{Arman Cohan}.} \bibinfo{year}{2020}\natexlab{}.
\newblock \showarticletitle{Longformer: The long-document transformer}.
\newblock \bibinfo{journal}{\emph{arXiv preprint arXiv:2004.05150}} (\bibinfo{year}{2020}).
\newblock


\bibitem[Bommasani et~al\mbox{.}(2021)]%
        {bommasani2021opportunities}
\bibfield{author}{\bibinfo{person}{Rishi Bommasani}, \bibinfo{person}{Drew~A Hudson}, \bibinfo{person}{Ehsan Adeli}, \bibinfo{person}{Russ Altman}, \bibinfo{person}{Simran Arora}, \bibinfo{person}{Sydney von Arx}, \bibinfo{person}{Michael~S Bernstein}, \bibinfo{person}{Jeannette Bohg}, \bibinfo{person}{Antoine Bosselut}, \bibinfo{person}{Emma Brunskill}, {et~al\mbox{.}}} \bibinfo{year}{2021}\natexlab{}.
\newblock \showarticletitle{On the opportunities and risks of foundation models}.
\newblock \bibinfo{journal}{\emph{arXiv preprint arXiv:2108.07258}} (\bibinfo{year}{2021}).
\newblock


\bibitem[Brown et~al\mbox{.}(2020)]%
        {brown2020language}
\bibfield{author}{\bibinfo{person}{Tom Brown}, \bibinfo{person}{Benjamin Mann}, \bibinfo{person}{Nick Ryder}, \bibinfo{person}{Melanie Subbiah}, \bibinfo{person}{Jared~D Kaplan}, \bibinfo{person}{Prafulla Dhariwal}, \bibinfo{person}{Arvind Neelakantan}, \bibinfo{person}{Pranav Shyam}, \bibinfo{person}{Girish Sastry}, \bibinfo{person}{Amanda Askell}, {et~al\mbox{.}}} \bibinfo{year}{2020}\natexlab{}.
\newblock \showarticletitle{Language models are few-shot learners}.
\newblock \bibinfo{journal}{\emph{Advances in neural information processing systems}}  \bibinfo{volume}{33} (\bibinfo{year}{2020}), \bibinfo{pages}{1877--1901}.
\newblock


\bibitem[Chen et~al\mbox{.}(2023b)]%
        {chen2023llm}
\bibfield{author}{\bibinfo{person}{Siyuan Chen}, \bibinfo{person}{Mengyue Wu}, \bibinfo{person}{Kenny~Q Zhu}, \bibinfo{person}{Kunyao Lan}, \bibinfo{person}{Zhiling Zhang}, {and} \bibinfo{person}{Lyuchun Cui}.} \bibinfo{year}{2023}\natexlab{b}.
\newblock \showarticletitle{LLM-empowered Chatbots for Psychiatrist and Patient Simulation: Application and Evaluation}.
\newblock \bibinfo{journal}{\emph{arXiv preprint arXiv:2305.13614}} (\bibinfo{year}{2023}).
\newblock


\bibitem[Chen et~al\mbox{.}(2023a)]%
        {chen2023dynamic_nm}
\bibfield{author}{\bibinfo{person}{Zhaodong Chen}, \bibinfo{person}{Zheng Qu}, \bibinfo{person}{Yuying Quan}, \bibinfo{person}{Liu Liu}, \bibinfo{person}{Yufei Ding}, {and} \bibinfo{person}{Yuan Xie}.} \bibinfo{year}{2023}\natexlab{a}.
\newblock \showarticletitle{Dynamic n: M fine-grained structured sparse attention mechanism}. In \bibinfo{booktitle}{\emph{Proceedings of the 28th ACM SIGPLAN Annual Symposium on Principles and Practice of Parallel Programming}}. \bibinfo{pages}{369--379}.
\newblock


\bibitem[Child et~al\mbox{.}(2019)]%
        {child2019spTrans}
\bibfield{author}{\bibinfo{person}{Rewon Child}, \bibinfo{person}{Scott Gray}, \bibinfo{person}{Alec Radford}, {and} \bibinfo{person}{Ilya Sutskever}.} \bibinfo{year}{2019}\natexlab{}.
\newblock \showarticletitle{Generating long sequences with sparse transformers}.
\newblock \bibinfo{journal}{\emph{arXiv preprint arXiv:1904.10509}} (\bibinfo{year}{2019}).
\newblock


\bibitem[Choquette et~al\mbox{.}(2021)]%
        {a100-2021}
\bibfield{author}{\bibinfo{person}{Jack Choquette}, \bibinfo{person}{Wishwesh Gandhi}, \bibinfo{person}{Olivier Giroux}, \bibinfo{person}{Nick Stam}, {and} \bibinfo{person}{Ronny Krashinsky}.} \bibinfo{year}{2021}\natexlab{}.
\newblock \showarticletitle{Nvidia a100 tensor core gpu: Performance and innovation}.
\newblock \bibinfo{journal}{\emph{IEEE Micro}} \bibinfo{volume}{41}, \bibinfo{number}{2} (\bibinfo{year}{2021}), \bibinfo{pages}{29--35}.
\newblock


\bibitem[Computer(2023)]%
        {together2023redpajama}
\bibfield{author}{\bibinfo{person}{Together Computer}.} \bibinfo{year}{2023}\natexlab{}.
\newblock \bibinfo{booktitle}{\emph{RedPajama: An Open Source Recipe to Reproduce LLaMA training dataset}}.
\newblock
\urldef\tempurl%
\url{https://github.com/togethercomputer/RedPajama-Data}
\showURL{%
\tempurl}


\bibitem[Cui et~al\mbox{.}(2023)]%
        {230616092}
\bibfield{author}{\bibinfo{person}{Jiaxi Cui}, \bibinfo{person}{Zongjian Li}, \bibinfo{person}{Yang Yan}, \bibinfo{person}{Bohua Chen}, {and} \bibinfo{person}{Li Yuan}.} \bibinfo{year}{2023}\natexlab{}.
\newblock \showarticletitle{ChatLaw: Open-Source Legal Large Language Model with Integrated External Knowledge Bases}.
\newblock \bibinfo{journal}{\emph{arXiv:2306.16092}} (\bibinfo{year}{2023}).
\newblock


\bibitem[Deng et~al\mbox{.}(2020)]%
        {deng2020model}
\bibfield{author}{\bibinfo{person}{Lei Deng}, \bibinfo{person}{Guoqi Li}, \bibinfo{person}{Song Han}, \bibinfo{person}{Luping Shi}, {and} \bibinfo{person}{Yuan Xie}.} \bibinfo{year}{2020}\natexlab{}.
\newblock \showarticletitle{Model compression and hardware acceleration for neural networks: A comprehensive survey}.
\newblock \bibinfo{journal}{\emph{Proc. IEEE}} \bibinfo{volume}{108}, \bibinfo{number}{4} (\bibinfo{year}{2020}), \bibinfo{pages}{485--532}.
\newblock


\bibitem[Dettmers et~al\mbox{.}(2022a)]%
        {llmint8}
\bibfield{author}{\bibinfo{person}{Tim Dettmers}, \bibinfo{person}{Mike Lewis}, \bibinfo{person}{Younes Belkada}, {and} \bibinfo{person}{Luke Zettlemoyer}.} \bibinfo{year}{2022}\natexlab{a}.
\newblock \showarticletitle{Llm. int8 (): 8-bit matrix multiplication for transformers at scale}.
\newblock \bibinfo{journal}{\emph{arXiv preprint arXiv:2208.07339}} (\bibinfo{year}{2022}).
\newblock


\bibitem[Dettmers et~al\mbox{.}(2022b)]%
        {dettmers2022llm_int8}
\bibfield{author}{\bibinfo{person}{Tim Dettmers}, \bibinfo{person}{Mike Lewis}, \bibinfo{person}{Younes Belkada}, {and} \bibinfo{person}{Luke Zettlemoyer}.} \bibinfo{year}{2022}\natexlab{b}.
\newblock \showarticletitle{Llm. int8 (): 8-bit matrix multiplication for transformers at scale}.
\newblock \bibinfo{journal}{\emph{arXiv preprint arXiv:2208.07339}} (\bibinfo{year}{2022}).
\newblock


\bibitem[Devlin et~al\mbox{.}(2019)]%
        {devlin2019bert}
\bibfield{author}{\bibinfo{person}{Jacob Devlin}, \bibinfo{person}{Ming-Wei Chang}, \bibinfo{person}{Kenton Lee}, {and} \bibinfo{person}{Kristina Toutanova}.} \bibinfo{year}{2019}\natexlab{}.
\newblock \bibinfo{title}{BERT: Pre-training of Deep Bidirectional Transformers for Language Understanding}.
\newblock
\newblock
\showeprint[arxiv]{1810.04805}~[cs.CL]


\bibitem[Fan et~al\mbox{.}(2022)]%
        {micro2022bufferfly}
\bibfield{author}{\bibinfo{person}{Hongxiang Fan}, \bibinfo{person}{Thomas Chau}, \bibinfo{person}{Stylianos~I. Venieris}, \bibinfo{person}{Royson Lee}, \bibinfo{person}{Alexandros Kouris}, \bibinfo{person}{Wayne Luk}, \bibinfo{person}{Nicholas~D. Lane}, {and} \bibinfo{person}{Mohamed~S. Abdelfattah}.} \bibinfo{year}{2022}\natexlab{}.
\newblock \showarticletitle{Adaptable Butterfly Accelerator for Attention-based NNs via Hardware and Algorithm Co-design}. In \bibinfo{booktitle}{\emph{2022 55th IEEE/ACM International Symposium on Microarchitecture (MICRO)}}. \bibinfo{pages}{599--615}.
\newblock
\urldef\tempurl%
\url{https://doi.org/10.1109/MICRO56248.2022.00050}
\showDOI{\tempurl}


\bibitem[Feng et~al\mbox{.}(2023)]%
        {feng2023diffuser}
\bibfield{author}{\bibinfo{person}{Aosong Feng}, \bibinfo{person}{Irene Li}, \bibinfo{person}{Yuang Jiang}, {and} \bibinfo{person}{Rex Ying}.} \bibinfo{year}{2023}\natexlab{}.
\newblock \showarticletitle{Diffuser: efficient transformers with multi-hop attention diffusion for long sequences}. In \bibinfo{booktitle}{\emph{Proceedings of the AAAI Conference on Artificial Intelligence}}, Vol.~\bibinfo{volume}{37}. \bibinfo{pages}{12772--12780}.
\newblock


\bibitem[Frantar and Alistarh(2023)]%
        {frantar2023sparsegpt}
\bibfield{author}{\bibinfo{person}{Elias Frantar} {and} \bibinfo{person}{Dan Alistarh}.} \bibinfo{year}{2023}\natexlab{}.
\newblock \showarticletitle{SparseGPT: Massive Language Models Can Be Accurately Pruned in One-Shot}.
\newblock  (\bibinfo{year}{2023}).
\newblock


\bibitem[Frantar et~al\mbox{.}(2022)]%
        {frantar2022gptq}
\bibfield{author}{\bibinfo{person}{Elias Frantar}, \bibinfo{person}{Saleh Ashkboos}, \bibinfo{person}{Torsten Hoefler}, {and} \bibinfo{person}{Dan Alistarh}.} \bibinfo{year}{2022}\natexlab{}.
\newblock \showarticletitle{Gptq: Accurate post-training quantization for generative pre-trained transformers}.
\newblock \bibinfo{journal}{\emph{arXiv preprint arXiv:2210.17323}} (\bibinfo{year}{2022}).
\newblock


\bibitem[Gong et~al\mbox{.}(2022)]%
        {gong2022n3h}
\bibfield{author}{\bibinfo{person}{Yu Gong}, \bibinfo{person}{Zhihan Xu}, \bibinfo{person}{Zhezhi He}, \bibinfo{person}{Weifeng Zhang}, \bibinfo{person}{Xiaobing Tu}, \bibinfo{person}{Xiaoyao Liang}, {and} \bibinfo{person}{Li Jiang}.} \bibinfo{year}{2022}\natexlab{}.
\newblock \showarticletitle{N3H-core: Neuron-designed neural network accelerator via FPGA-based heterogeneous computing cores}. In \bibinfo{booktitle}{\emph{Proceedings of the 2022 ACM/SIGDA International Symposium on Field-Programmable Gate Arrays}}. \bibinfo{pages}{112--122}.
\newblock


\bibitem[Guo et~al\mbox{.}(2017)]%
        {guo2017angel}
\bibfield{author}{\bibinfo{person}{Kaiyuan Guo}, \bibinfo{person}{Lingzhi Sui}, \bibinfo{person}{Jiantao Qiu}, \bibinfo{person}{Jincheng Yu}, \bibinfo{person}{Junbin Wang}, \bibinfo{person}{Song Yao}, \bibinfo{person}{Song Han}, \bibinfo{person}{Yu Wang}, {and} \bibinfo{person}{Huazhong Yang}.} \bibinfo{year}{2017}\natexlab{}.
\newblock \showarticletitle{Angel-eye: A complete design flow for mapping CNN onto embedded FPGA}.
\newblock \bibinfo{journal}{\emph{IEEE transactions on computer-aided design of integrated circuits and systems}} \bibinfo{volume}{37}, \bibinfo{number}{1} (\bibinfo{year}{2017}), \bibinfo{pages}{35--47}.
\newblock


\bibitem[Ham et~al\mbox{.}(2020)]%
        {ham2020a3}
\bibfield{author}{\bibinfo{person}{Tae~Jun Ham}, \bibinfo{person}{Sung~Jun Jung}, \bibinfo{person}{Seonghak Kim}, \bibinfo{person}{Young~H. Oh}, \bibinfo{person}{Yeonhong Park}, \bibinfo{person}{Yoonho Song}, \bibinfo{person}{Jung-Hun Park}, \bibinfo{person}{Sanghee Lee}, \bibinfo{person}{Kyoung Park}, \bibinfo{person}{Jae~W. Lee}, {and} \bibinfo{person}{Deog-Kyoon Jeong}.} \bibinfo{year}{2020}\natexlab{}.
\newblock \bibinfo{title}{A$^3$: Accelerating Attention Mechanisms in Neural Networks with Approximation}.
\newblock
\newblock
\showeprint[arxiv]{2002.10941}~[cs.DC]


\bibitem[Ham et~al\mbox{.}(2021)]%
        {isca2021elsa}
\bibfield{author}{\bibinfo{person}{Tae~Jun Ham}, \bibinfo{person}{Yejin Lee}, \bibinfo{person}{Seong~Hoon Seo}, \bibinfo{person}{Soosung Kim}, \bibinfo{person}{Hyunji Choi}, \bibinfo{person}{Sung~Jun Jung}, {and} \bibinfo{person}{Jae~W. Lee}.} \bibinfo{year}{2021}\natexlab{}.
\newblock \showarticletitle{ELSA: Hardware-Software Co-design for Efficient, Lightweight Self-Attention Mechanism in Neural Networks}. In \bibinfo{booktitle}{\emph{2021 ACM/IEEE 48th Annual International Symposium on Computer Architecture (ISCA)}}. \bibinfo{pages}{692--705}.
\newblock
\urldef\tempurl%
\url{https://doi.org/10.1109/ISCA52012.2021.00060}
\showDOI{\tempurl}


\bibitem[Hong et~al\mbox{.}(2022)]%
        {micro2022DFX}
\bibfield{author}{\bibinfo{person}{Seongmin Hong}, \bibinfo{person}{Seungjae Moon}, \bibinfo{person}{Junsoo Kim}, \bibinfo{person}{Sungjae Lee}, \bibinfo{person}{Minsub Kim}, \bibinfo{person}{Dongsoo Lee}, {and} \bibinfo{person}{Joo-Young Kim}.} \bibinfo{year}{2022}\natexlab{}.
\newblock \showarticletitle{DFX: A Low-latency Multi-FPGA Appliance for Accelerating Transformer-based Text Generation}. In \bibinfo{booktitle}{\emph{2022 IEEE Hot Chips 34 Symposium (HCS)}}. \bibinfo{pages}{1--17}.
\newblock
\urldef\tempurl%
\url{https://doi.org/10.1109/HCS55958.2022.9895626}
\showDOI{\tempurl}


\bibitem[Jeong(2023)]%
        {jeong2023study}
\bibfield{author}{\bibinfo{person}{Cheonsu Jeong}.} \bibinfo{year}{2023}\natexlab{}.
\newblock \showarticletitle{A Study on the Implementation of Generative AI Services Using an Enterprise Data-Based LLM Application Architecture}.
\newblock \bibinfo{journal}{\emph{arXiv preprint arXiv:2309.01105}} (\bibinfo{year}{2023}).
\newblock


\bibitem[Kao et~al\mbox{.}(2023)]%
        {hpca23flat}
\bibfield{author}{\bibinfo{person}{Sheng-Chun Kao}, \bibinfo{person}{Suvinay Subramanian}, \bibinfo{person}{Gaurav Agrawal}, \bibinfo{person}{Amir Yazdanbakhsh}, {and} \bibinfo{person}{Tushar Krishna}.} \bibinfo{year}{2023}\natexlab{}.
\newblock \showarticletitle{FLAT: An Optimized Dataflow for Mitigating Attention Bottlenecks}. In \bibinfo{booktitle}{\emph{Proceedings of the 28th ACM International Conference on Architectural Support for Programming Languages and Operating Systems, Volume 2}} (Vancouver, BC, Canada) \emph{(\bibinfo{series}{ASPLOS 2023})}. \bibinfo{publisher}{Association for Computing Machinery}, \bibinfo{address}{New York, NY, USA}, \bibinfo{pages}{295–310}.
\newblock
\showISBNx{9781450399166}
\urldef\tempurl%
\url{https://doi.org/10.1145/3575693.3575747}
\showDOI{\tempurl}


\bibitem[Kim et~al\mbox{.}(2023)]%
        {squeezellm}
\bibfield{author}{\bibinfo{person}{Sehoon Kim}, \bibinfo{person}{Coleman Hooper}, \bibinfo{person}{Amir Gholami}, \bibinfo{person}{Zhen Dong}, \bibinfo{person}{Xiuyu Li}, \bibinfo{person}{Sheng Shen}, \bibinfo{person}{Michael~W Mahoney}, {and} \bibinfo{person}{Kurt Keutzer}.} \bibinfo{year}{2023}\natexlab{}.
\newblock \showarticletitle{SqueezeLLM: Dense-and-Sparse Quantization}.
\newblock \bibinfo{journal}{\emph{arXiv preprint arXiv:2306.07629}} (\bibinfo{year}{2023}).
\newblock


\bibitem[Kitaev et~al\mbox{.}(2020)]%
        {kitaev2020reformer}
\bibfield{author}{\bibinfo{person}{Nikita Kitaev}, \bibinfo{person}{{\L}ukasz Kaiser}, {and} \bibinfo{person}{Anselm Levskaya}.} \bibinfo{year}{2020}\natexlab{}.
\newblock \showarticletitle{Reformer: The efficient transformer}.
\newblock \bibinfo{journal}{\emph{arXiv preprint arXiv:2001.04451}} (\bibinfo{year}{2020}).
\newblock


\bibitem[Kwon et~al\mbox{.}(2022)]%
        {kwon2022fast}
\bibfield{author}{\bibinfo{person}{Woosuk Kwon}, \bibinfo{person}{Sehoon Kim}, \bibinfo{person}{Michael~W Mahoney}, \bibinfo{person}{Joseph Hassoun}, \bibinfo{person}{Kurt Keutzer}, {and} \bibinfo{person}{Amir Gholami}.} \bibinfo{year}{2022}\natexlab{}.
\newblock \showarticletitle{A fast post-training pruning framework for transformers}.
\newblock \bibinfo{journal}{\emph{Advances in Neural Information Processing Systems}}  \bibinfo{volume}{35} (\bibinfo{year}{2022}), \bibinfo{pages}{24101--24116}.
\newblock


\bibitem[Kwon et~al\mbox{.}(2023)]%
        {kwon2023efficient}
\bibfield{author}{\bibinfo{person}{Woosuk Kwon}, \bibinfo{person}{Zhuohan Li}, \bibinfo{person}{Siyuan Zhuang}, \bibinfo{person}{Ying Sheng}, \bibinfo{person}{Lianmin Zheng}, \bibinfo{person}{Cody~Hao Yu}, \bibinfo{person}{Joseph Gonzalez}, \bibinfo{person}{Hao Zhang}, {and} \bibinfo{person}{Ion Stoica}.} \bibinfo{year}{2023}\natexlab{}.
\newblock \showarticletitle{Efficient Memory Management for Large Language Model Serving with PagedAttention}. In \bibinfo{booktitle}{\emph{Proceedings of the 29th Symposium on Operating Systems Principles}}. \bibinfo{pages}{611--626}.
\newblock


\bibitem[Li et~al\mbox{.}(2020)]%
        {li2020ftrans}
\bibfield{author}{\bibinfo{person}{Bingbing Li}, \bibinfo{person}{Santosh Pandey}, \bibinfo{person}{Haowen Fang}, \bibinfo{person}{Yanjun Lyv}, \bibinfo{person}{Ji Li}, \bibinfo{person}{Jieyang Chen}, \bibinfo{person}{Mimi Xie}, \bibinfo{person}{Lipeng Wan}, \bibinfo{person}{Hang Liu}, {and} \bibinfo{person}{Caiwen Ding}.} \bibinfo{year}{2020}\natexlab{}.
\newblock \bibinfo{title}{FTRANS: Energy-Efficient Acceleration of Transformers using FPGA}.
\newblock
\newblock
\showeprint[arxiv]{2007.08563}~[cs.DC]


\bibitem[Lin et~al\mbox{.}(2023)]%
        {lin2023awq}
\bibfield{author}{\bibinfo{person}{Ji Lin}, \bibinfo{person}{Jiaming Tang}, \bibinfo{person}{Haotian Tang}, \bibinfo{person}{Shang Yang}, \bibinfo{person}{Xingyu Dang}, {and} \bibinfo{person}{Song Han}.} \bibinfo{year}{2023}\natexlab{}.
\newblock \showarticletitle{AWQ: Activation-aware Weight Quantization for LLM Compression and Acceleration}.
\newblock \bibinfo{journal}{\emph{arXiv preprint arXiv:2306.00978}} (\bibinfo{year}{2023}).
\newblock


\bibitem[Lu et~al\mbox{.}(2021)]%
        {micro2021sanger}
\bibfield{author}{\bibinfo{person}{Liqiang Lu}, \bibinfo{person}{Yicheng Jin}, \bibinfo{person}{Hangrui Bi}, \bibinfo{person}{Zizhang Luo}, \bibinfo{person}{Peng Li}, \bibinfo{person}{Tao Wang}, {and} \bibinfo{person}{Yun Liang}.} \bibinfo{year}{2021}\natexlab{}.
\newblock \showarticletitle{Sanger: A Co-Design Framework for Enabling Sparse Attention Using Reconfigurable Architecture}. In \bibinfo{booktitle}{\emph{MICRO-54: 54th Annual IEEE/ACM International Symposium on Microarchitecture}} (Virtual Event, Greece) \emph{(\bibinfo{series}{MICRO '21})}. \bibinfo{publisher}{Association for Computing Machinery}, \bibinfo{address}{New York, NY, USA}, \bibinfo{pages}{977–991}.
\newblock
\showISBNx{9781450385572}
\urldef\tempurl%
\url{https://doi.org/10.1145/3466752.3480125}
\showDOI{\tempurl}


\bibitem[Merity et~al\mbox{.}(2016)]%
        {merity2016wikitext2}
\bibfield{author}{\bibinfo{person}{Stephen Merity}, \bibinfo{person}{Caiming Xiong}, \bibinfo{person}{James Bradbury}, {and} \bibinfo{person}{Richard Socher}.} \bibinfo{year}{2016}\natexlab{}.
\newblock \showarticletitle{Pointer sentinel mixture models}.
\newblock \bibinfo{journal}{\emph{arXiv preprint arXiv:1609.07843}} (\bibinfo{year}{2016}).
\newblock


\bibitem[Naveed et~al\mbox{.}(2023)]%
        {Naveed2023ACO}
\bibfield{author}{\bibinfo{person}{Humza Naveed}, \bibinfo{person}{Asad~Ullah Khan}, \bibinfo{person}{Shi Qiu}, \bibinfo{person}{Muhammad Saqib}, \bibinfo{person}{Saeed Anwar}, \bibinfo{person}{Muhammad Usman}, \bibinfo{person}{Nick Barnes}, {and} \bibinfo{person}{Ajmal~S. Mian}.} \bibinfo{year}{2023}\natexlab{}.
\newblock \showarticletitle{A Comprehensive Overview of Large Language Models}.
\newblock \bibinfo{journal}{\emph{ArXiv}}  \bibinfo{volume}{abs/2307.06435} (\bibinfo{year}{2023}).
\newblock


\bibitem[Qin et~al\mbox{.}(2023)]%
        {isca2023fact}
\bibfield{author}{\bibinfo{person}{Yubin Qin}, \bibinfo{person}{Yang Wang}, \bibinfo{person}{Dazheng Deng}, \bibinfo{person}{Zhiren Zhao}, \bibinfo{person}{Xiaolong Yang}, \bibinfo{person}{Leibo Liu}, \bibinfo{person}{Shaojun Wei}, \bibinfo{person}{Yang Hu}, {and} \bibinfo{person}{Shouyi Yin}.} \bibinfo{year}{2023}\natexlab{}.
\newblock \showarticletitle{FACT: FFN-Attention Co-Optimized Transformer Architecture with Eager Correlation Prediction}. In \bibinfo{booktitle}{\emph{Proceedings of the 50th Annual International Symposium on Computer Architecture}} (Orlando, FL, USA) \emph{(\bibinfo{series}{ISCA '23})}. \bibinfo{publisher}{Association for Computing Machinery}, \bibinfo{address}{New York, NY, USA}, Article \bibinfo{articleno}{22}, \bibinfo{numpages}{14}~pages.
\newblock
\showISBNx{9798400700958}
\urldef\tempurl%
\url{https://doi.org/10.1145/3579371.3589057}
\showDOI{\tempurl}


\bibitem[Srivastava et~al\mbox{.}(2020)]%
        {matraptor-2020}
\bibfield{author}{\bibinfo{person}{Nitish Srivastava}, \bibinfo{person}{Hanchen Jin}, \bibinfo{person}{Jie Liu}, \bibinfo{person}{David Albonesi}, {and} \bibinfo{person}{Zhiru Zhang}.} \bibinfo{year}{2020}\natexlab{}.
\newblock \showarticletitle{Matraptor: A sparse-sparse matrix multiplication accelerator based on row-wise product}. In \bibinfo{booktitle}{\emph{2020 53rd Annual IEEE/ACM International Symposium on Microarchitecture (MICRO)}}. IEEE, \bibinfo{pages}{766--780}.
\newblock


\bibitem[Sun et~al\mbox{.}(2022)]%
        {sun2022film}
\bibfield{author}{\bibinfo{person}{Mengshu Sun}, \bibinfo{person}{Zhengang Li}, \bibinfo{person}{Alec Lu}, \bibinfo{person}{Yanyu Li}, \bibinfo{person}{Sung-En Chang}, \bibinfo{person}{Xiaolong Ma}, \bibinfo{person}{Xue Lin}, {and} \bibinfo{person}{Zhenman Fang}.} \bibinfo{year}{2022}\natexlab{}.
\newblock \showarticletitle{Film-qnn: Efficient fpga acceleration of deep neural networks with intra-layer, mixed-precision quantization}. In \bibinfo{booktitle}{\emph{Proceedings of the 2022 ACM/SIGDA International Symposium on Field-Programmable Gate Arrays}}. \bibinfo{pages}{134--145}.
\newblock


\bibitem[Thirunavukarasu et~al\mbox{.}(2023)]%
        {thirunavukarasu2023large}
\bibfield{author}{\bibinfo{person}{Arun~James Thirunavukarasu}, \bibinfo{person}{Darren Shu~Jeng Ting}, \bibinfo{person}{Kabilan Elangovan}, \bibinfo{person}{Laura Gutierrez}, \bibinfo{person}{Ting~Fang Tan}, {and} \bibinfo{person}{Daniel Shu~Wei Ting}.} \bibinfo{year}{2023}\natexlab{}.
\newblock \showarticletitle{Large language models in medicine}.
\newblock \bibinfo{journal}{\emph{Nature medicine}} \bibinfo{volume}{29}, \bibinfo{number}{8} (\bibinfo{year}{2023}), \bibinfo{pages}{1930--1940}.
\newblock


\bibitem[Touvron et~al\mbox{.}(2023)]%
        {touvron2023llama2}
\bibfield{author}{\bibinfo{person}{Hugo Touvron}, \bibinfo{person}{Thibaut Lavril}, \bibinfo{person}{Gautier Izacard}, \bibinfo{person}{Xavier Martinet}, \bibinfo{person}{Marie-Anne Lachaux}, \bibinfo{person}{Timoth{\'e}e Lacroix}, \bibinfo{person}{Baptiste Rozi{\`e}re}, \bibinfo{person}{Naman Goyal}, \bibinfo{person}{Eric Hambro}, \bibinfo{person}{Faisal Azhar}, {et~al\mbox{.}}} \bibinfo{year}{2023}\natexlab{}.
\newblock \showarticletitle{Llama: Open and efficient foundation language models}.
\newblock \bibinfo{journal}{\emph{arXiv preprint arXiv:2302.13971}} (\bibinfo{year}{2023}).
\newblock


\bibitem[Wang et~al\mbox{.}(2023a)]%
        {Wang2023PractitionersEO}
\bibfield{author}{\bibinfo{person}{Chaozheng Wang}, \bibinfo{person}{Junhao Hu}, \bibinfo{person}{Cuiyun Gao}, \bibinfo{person}{Yu Jin}, \bibinfo{person}{Tao Xie}, \bibinfo{person}{Hailiang Huang}, \bibinfo{person}{Zhenyu Lei}, {and} \bibinfo{person}{Yuetang Deng}.} \bibinfo{year}{2023}\natexlab{a}.
\newblock \showarticletitle{Practitioners' Expectations on Code Completion}.
\newblock \bibinfo{journal}{\emph{ArXiv}}  \bibinfo{volume}{abs/2301.03846} (\bibinfo{year}{2023}).
\newblock


\bibitem[Wang et~al\mbox{.}(2022)]%
        {wang2022logic}
\bibfield{author}{\bibinfo{person}{Erwei Wang}, \bibinfo{person}{James~J Davis}, \bibinfo{person}{Georgios-Ilias Stavrou}, \bibinfo{person}{Peter~YK Cheung}, \bibinfo{person}{George~A Constantinides}, {and} \bibinfo{person}{Mohamed Abdelfattah}.} \bibinfo{year}{2022}\natexlab{}.
\newblock \showarticletitle{Logic shrinkage: Learned FPGA netlist sparsity for efficient neural network inference}. In \bibinfo{booktitle}{\emph{Proceedings of the 2022 ACM/SIGDA International Symposium on Field-Programmable Gate Arrays}}. \bibinfo{pages}{101--111}.
\newblock


\bibitem[Wang et~al\mbox{.}(2023c)]%
        {CTA2023}
\bibfield{author}{\bibinfo{person}{Haoran Wang}, \bibinfo{person}{Haobo Xu}, \bibinfo{person}{Ying Wang}, {and} \bibinfo{person}{Yinhe Han}.} \bibinfo{year}{2023}\natexlab{c}.
\newblock \showarticletitle{CTA: Hardware-Software Co-design for Compressed Token Attention Mechanism}. In \bibinfo{booktitle}{\emph{2023 IEEE International Symposium on High-Performance Computer Architecture (HPCA)}}. \bibinfo{pages}{429--441}.
\newblock
\urldef\tempurl%
\url{https://doi.org/10.1109/HPCA56546.2023.10070997}
\showDOI{\tempurl}


\bibitem[Wang et~al\mbox{.}(2021)]%
        {wang2021spatten}
\bibfield{author}{\bibinfo{person}{Hanrui Wang}, \bibinfo{person}{Zhekai Zhang}, {and} \bibinfo{person}{Song Han}.} \bibinfo{year}{2021}\natexlab{}.
\newblock \showarticletitle{Spatten: Efficient sparse attention architecture with cascade token and head pruning}. In \bibinfo{booktitle}{\emph{2021 IEEE International Symposium on High-Performance Computer Architecture (HPCA)}}. IEEE, \bibinfo{pages}{97--110}.
\newblock


\bibitem[Wang et~al\mbox{.}(2020)]%
        {wang2020shuhai}
\bibfield{author}{\bibinfo{person}{Zeke Wang}, \bibinfo{person}{Hongjing Huang}, \bibinfo{person}{Jie Zhang}, {and} \bibinfo{person}{Gustavo Alonso}.} \bibinfo{year}{2020}\natexlab{}.
\newblock \showarticletitle{Shuhai: Benchmarking high bandwidth memory on fpgas}. In \bibinfo{booktitle}{\emph{2020 IEEE 28th Annual International Symposium on Field-Programmable Custom Computing Machines (FCCM)}}. IEEE, \bibinfo{pages}{111--119}.
\newblock


\bibitem[Wang et~al\mbox{.}(2023b)]%
        {wang2023cosa}
\bibfield{author}{\bibinfo{person}{Zhican Wang}, \bibinfo{person}{Gang Wang}, \bibinfo{person}{Honglan Jiang}, \bibinfo{person}{Ningyi Xu}, {and} \bibinfo{person}{Guanghui He}.} \bibinfo{year}{2023}\natexlab{b}.
\newblock \showarticletitle{COSA: Co-Operative Systolic Arrays for Multi-head Attention Mechanism in Neural Network using Hybrid Data Reuse and Fusion Methodologies}. In \bibinfo{booktitle}{\emph{2023 60th ACM/IEEE Design Automation Conference (DAC)}}. IEEE, \bibinfo{pages}{1--6}.
\newblock


\bibitem[Wei et~al\mbox{.}(2022)]%
        {wei2022emergent}
\bibfield{author}{\bibinfo{person}{Jason Wei}, \bibinfo{person}{Yi Tay}, \bibinfo{person}{Rishi Bommasani}, \bibinfo{person}{Colin Raffel}, \bibinfo{person}{Barret Zoph}, \bibinfo{person}{Sebastian Borgeaud}, \bibinfo{person}{Dani Yogatama}, \bibinfo{person}{Maarten Bosma}, \bibinfo{person}{Denny Zhou}, \bibinfo{person}{Donald Metzler}, {et~al\mbox{.}}} \bibinfo{year}{2022}\natexlab{}.
\newblock \showarticletitle{Emergent abilities of large language models}.
\newblock \bibinfo{journal}{\emph{arXiv preprint arXiv:2206.07682}} (\bibinfo{year}{2022}).
\newblock


\bibitem[Xiao et~al\mbox{.}(2023)]%
        {xiao2023smoothquant}
\bibfield{author}{\bibinfo{person}{Guangxuan Xiao}, \bibinfo{person}{Ji Lin}, \bibinfo{person}{Mickael Seznec}, \bibinfo{person}{Hao Wu}, \bibinfo{person}{Julien Demouth}, {and} \bibinfo{person}{Song Han}.} \bibinfo{year}{2023}\natexlab{}.
\newblock \showarticletitle{Smoothquant: Accurate and efficient post-training quantization for large language models}. In \bibinfo{booktitle}{\emph{International Conference on Machine Learning}}. PMLR, \bibinfo{pages}{38087--38099}.
\newblock


\bibitem[Xilinx(2021)]%
        {fpgaU280}
\bibfield{author}{\bibinfo{person}{Xilinx}.} \bibinfo{year}{2021}\natexlab{}.
\newblock \bibinfo{title}{Alveo U280 Data Center Accelerator Card Data Sheet}.
\newblock \bibinfo{howpublished}{\url{https://docs.xilinx.com/v/u/en-US/ds963-u280}}.
\newblock


\bibitem[Xilinx(2023)]%
        {VersalVH1582}
\bibfield{author}{\bibinfo{person}{Xilinx}.} \bibinfo{year}{2023}\natexlab{}.
\newblock \bibinfo{title}{Versal™ Architecture and Product Data Sheet}.
\newblock \bibinfo{howpublished}{\url{https://docs.xilinx.com/v/u/en-US/ds950-versal-overview}}.
\newblock


\bibitem[Yao et~al\mbox{.}(2022)]%
        {yao2022zeroquant}
\bibfield{author}{\bibinfo{person}{Zhewei Yao}, \bibinfo{person}{Reza Yazdani~Aminabadi}, \bibinfo{person}{Minjia Zhang}, \bibinfo{person}{Xiaoxia Wu}, \bibinfo{person}{Conglong Li}, {and} \bibinfo{person}{Yuxiong He}.} \bibinfo{year}{2022}\natexlab{}.
\newblock \showarticletitle{Zeroquant: Efficient and affordable post-training quantization for large-scale transformers}.
\newblock \bibinfo{journal}{\emph{Advances in Neural Information Processing Systems}}  \bibinfo{volume}{35} (\bibinfo{year}{2022}), \bibinfo{pages}{27168--27183}.
\newblock


\bibitem[Zaheer et~al\mbox{.}(2020)]%
        {zaheer2020bigbird}
\bibfield{author}{\bibinfo{person}{Manzil Zaheer}, \bibinfo{person}{Guru Guruganesh}, \bibinfo{person}{Kumar~Avinava Dubey}, \bibinfo{person}{Joshua Ainslie}, \bibinfo{person}{Chris Alberti}, \bibinfo{person}{Santiago Ontanon}, \bibinfo{person}{Philip Pham}, \bibinfo{person}{Anirudh Ravula}, \bibinfo{person}{Qifan Wang}, \bibinfo{person}{Li Yang}, {et~al\mbox{.}}} \bibinfo{year}{2020}\natexlab{}.
\newblock \showarticletitle{Big bird: Transformers for longer sequences}.
\newblock \bibinfo{journal}{\emph{Advances in neural information processing systems}}  \bibinfo{volume}{33} (\bibinfo{year}{2020}), \bibinfo{pages}{17283--17297}.
\newblock


\bibitem[Zhang et~al\mbox{.}({[n.\,d.]})]%
        {zhang2205opt}
\bibfield{author}{\bibinfo{person}{Susan Zhang}, \bibinfo{person}{Stephen Roller}, \bibinfo{person}{Naman Goyal}, \bibinfo{person}{Mikel Artetxe}, \bibinfo{person}{Moya Chen}, \bibinfo{person}{Shuohui Chen}, \bibinfo{person}{Christopher Dewan}, \bibinfo{person}{Mona Diab}, \bibinfo{person}{Xian Li}, \bibinfo{person}{Xi~Victoria Lin}, {et~al\mbox{.}}} \bibinfo{year}{[n.\,d.]}\natexlab{}.
\newblock \showarticletitle{Opt: Open pre-trained transformer language models, 2022}.
\newblock \bibinfo{journal}{\emph{URL https://arxiv. org/abs/2205.01068}} (\bibinfo{year}{[n.\,d.]}).
\newblock


\bibitem[Zhang et~al\mbox{.}(2021)]%
        {zhang2021fracbnn}
\bibfield{author}{\bibinfo{person}{Yichi Zhang}, \bibinfo{person}{Junhao Pan}, \bibinfo{person}{Xinheng Liu}, \bibinfo{person}{Hongzheng Chen}, \bibinfo{person}{Deming Chen}, {and} \bibinfo{person}{Zhiru Zhang}.} \bibinfo{year}{2021}\natexlab{}.
\newblock \showarticletitle{FracBNN: Accurate and FPGA-efficient binary neural networks with fractional activations}. In \bibinfo{booktitle}{\emph{The 2021 ACM/SIGDA International Symposium on Field-Programmable Gate Arrays}}. \bibinfo{pages}{171--182}.
\newblock


\bibitem[Zhao et~al\mbox{.}(2018)]%
        {zhao2018bandwidth}
\bibfield{author}{\bibinfo{person}{Han Zhao}, \bibinfo{person}{Quan Chen}, \bibinfo{person}{Yuxian Qiu}, \bibinfo{person}{Ming Wu}, \bibinfo{person}{Yao Shen}, \bibinfo{person}{Jingwen Leng}, \bibinfo{person}{Chao Li}, {and} \bibinfo{person}{Minyi Guo}.} \bibinfo{year}{2018}\natexlab{}.
\newblock \showarticletitle{Bandwidth and locality aware task-stealing for manycore architectures with bandwidth-asymmetric memory}.
\newblock \bibinfo{journal}{\emph{ACM Transactions on Architecture and Code Optimization (TACO)}} \bibinfo{volume}{15}, \bibinfo{number}{4} (\bibinfo{year}{2018}), \bibinfo{pages}{1--26}.
\newblock


\bibitem[Zhou et~al\mbox{.}(2021)]%
        {nm-sparse-arbitrary}
\bibfield{author}{\bibinfo{person}{Aojun Zhou}, \bibinfo{person}{Yukun Ma}, \bibinfo{person}{Junnan Zhu}, \bibinfo{person}{Jianbo Liu}, \bibinfo{person}{Zhijie Zhang}, \bibinfo{person}{Kun Yuan}, \bibinfo{person}{Wenxiu Sun}, {and} \bibinfo{person}{Hongsheng Li}.} \bibinfo{year}{2021}\natexlab{}.
\newblock \showarticletitle{Learning n: m fine-grained structured sparse neural networks from scratch}.
\newblock \bibinfo{journal}{\emph{arXiv preprint arXiv:2102.04010}} (\bibinfo{year}{2021}).
\newblock


\end{thebibliography}










\end{document}